\documentclass[prc,amsmath,superscriptaddress,showkeys,multicol,twocolumn]{revtex4}
\usepackage{gensymb}
\usepackage{graphicx,xcolor}
\usepackage{amssymb}
\usepackage{enumerate}
\usepackage{verbatim}
\usepackage{bm}
\usepackage{natbib}
\usepackage{float}
\usepackage{physics}
\usepackage{multirow}
\usepackage[outdir=./]{epstopdf}

\newcommand{\nc}{\newcommand}
\nc{\RR}[1]{\textcolor{red}{#1}}
\nc{\BB}[1]{\textcolor{blue}{#1}}
\nc{\GG}[1]{\textcolor{green}{#1}}
\nc{\PP}[1]{\textcolor{purple}{#1}}  

\nc{\Lhat}{\hat{L}}

\begin{document}
\title{Comparing event generator predictions and ab-initio calculations of $\nu$-$^{12}$C neutral current quasi-elastic scattering at 1 GeV}

\author{G.~B.~King}
\affiliation{Department of Physics, Washington University in St. Louis, MO 63130, USA}
\affiliation{Department of Physics and Astronomy, Michigan State University, East Lansing, MI 48824-1321}
\author{K.~Mahn}
\affiliation{Department of Physics and Astronomy, Michigan State University, East Lansing, MI 48824-1321}
\author{L.~Pickering}
\affiliation{Department of Physics and Astronomy, Michigan State University, East Lansing, MI 48824-1321}
\author{N.~Rocco}
\affiliation{Physics Division, Argonne National Laboratory, Argonne, Illinois, 60439 USA}
\affiliation{Theoretical Physics Department, Fermi National Laboratory, P.O. Box 500, Batavia, Illinois 60510, USA}

\date{\today}

\begin{abstract}
\begin{description}
\item[Background:]  The measurement of neutrino oscillations and exotic physics searches are important parts of the physics program in the near future, with new state-of-the-art experiments planned within the next decade. Future and modern experiments in these fields will make use of nuclear targets. Event Generators (EGs) are software used in the analysis of neutrino oscillation experiments. EGs use to predict kinematic observables for a range of neutrino energies. These simulations may lack physics captured by more rigorous theoretical calculations. 
\item[Purpose:] This work compares EG performance to nuclear theory calculations by comparing observables generated in the two frameworks. We provide a common set of definitions between theory and experiment and assess the physics contained in EG simulations. 
\item[Method:] Neutral current quasi-elastic (NCQE) scattering events for neutrinos and anti-neutrinos on a $^{12}$C target are simulated with a specific EG, NEUT, used by the T2K experiment for its analysis. The simulated cross sections are compared to analytic calculations from nuclear theory within the factorization scheme. 
We compare the NEUT implementation of two different models on nuclear spectral functions: the Relativistic Fermi Gas (RFG) and the correlated basis spectral function (CBF) to analytic calculations of the same models in the factorization scheme. For both nuclear physics models, we compare the appearance of features in the distributions relevant to experimental analyses.
\item[Results:] The peak of the cross section $d\sigma/ d\omega d\Omega$ is consistent in energy transfer, $\omega$, for RFG and CBF simulations. Qualitatively, the shape of the simulated distribution is similar to the one obtained through theory calculations; however, there are some discrepancies between the theory calculations and the NEUT simulation.
\item[Conclusions:] While the EG simulations and analytic calculations with the same model of nuclear dynamics show similar overall features, there are still differences between the two. These results demonstrate the importance of benchmarking EGs so their physics can be improved for analysis in future experiments.
\end{description}
\end{abstract}

\keywords{Event generators, neutrino oscillation, neutrino-nuclear interactions}

\maketitle


\section{Introduction}
\label{intro}

The measurement of neutrino oscillation parameters is one of the major goals of particle physics. Current \cite{t2k, nova, microboone, minerva} and future \cite{dune, hyperk} accelerator experiments measuring these parameters use nuclear targets, which makes having an adequate theory of neutrino-nucleus interactions crucial for the interpretation of data. Event generators (EGs) are the part of the experimental software that take in theoretical nuclear physics models and simulate the scattering process. In oscillation experiments, the energy of the incident neutrino is not known. The outgoing observables are related to the neutrino energy via the EG. Therefore, features of the differential cross section, such as the peak position and strength in different angles, are important assumptions tested by neutrino scattering experiments and informed by theoretical models. These types of issues are further discussed in \cite{nustec}. Empirical or fundamental parameters alter the strength of the cross section in the differential quantities. Uncertainties on the interaction model are significant in current experimental results \cite{t2k,nova}. In addition to neutrino oscillation experiments, models of neutral currents in EGs are important for the interpretation of data in exotic physics searches \cite{minos2011,nova2017,t2k2019}.

While EGs are built using the best phenomenological models available to cover the range of energies and scope of data being measured, approximations and inconsistent choices can still be made. Typically, approximate models like the Relativistic Fermi Gas (RFG) are used to model the dynamics of bound nucleons. Recently, the EG NEUT \cite{neut}, used by the T2K experiment for its analysis, has made an effort to include a more realistic spectral function for nucleons. The correlated basis spectral function (CBF) defined by Benhar {\it et al.} \cite{benhar1994}, commonly referred to in the EG community as simply the Spectral Function or SF model, was incorporated into NEUT \cite{furmanski2015} and is available to simulate events.  In addition to NEUT, there are other EGs on the market available for the simulation of neutrino interaction observables; however, for proof of concept, this work only compares results obtained with NEUT to analytic calculations of the differential cross section.

Making direct comparisons between EGs and state-of-the-art nuclear theory calculations can show how reasonable EG approximations are and where simulations can improve. Recent work presented in \cite{lovato2018} using Green's function Monte Carlo (GFMC) methods \cite{carlson2015} and a first principles description of nuclear dynamics \cite{wiringa1995, pieper2001} obtained results for a theoretical calculation of $\nu$-$^{12}$C neutral-current inclusive quasi-elastic scattering. 
Although sophisticated approaches such as these can calculate relevant observables, it is not currently practical to incorporate them into the neutrino oscillation analysis framework. EGs will thus play an important role in the future of these experiments and it is therefore important to understand how they compare to the best calculations available. It is through this sort of comparison that the EG community will understand the physics missing in these approximations and reliably estimate uncertainties for future neutrino oscillation analyses.

In order to compare with state-of-the-art calculations,
it is first necessary to establish common definitions of neutrino interactions between the experimental and theoretical communities. Furthermore, it is important to verify that the results obtained with EG simulations are consistent with theoretical predictions using the same physics.  Recent work has been published by the authors of \cite{rocco2019} analyzing the results of neutral- and charged-current neutrino-nucleus cross sections within the factorization scheme and using two state-of-the-art models of nuclear spectral functions. This same scheme can be employed using the spectral function for the RFG model of nucleon dynamics, as well. The consistency of EG predictions of neutral current quasi-elastic (NCQE) scattering observables for the RFG and CBF models can thus be compared to theoretical calculations. 
For this evaluation of EG performance, NCQE events for neutrinos and anti-neutrino scattering on a $^{12}$C target at a fixed beam energy of 1 GeV are simulated with NEUT v5.3.6 with modified features. Table \ref{tab:features} summarizes the modified features used in this work. The simulated events are used to obtain a double differential cross section that we compare with the prediction of the same observable obtained from theoretical calculations. In Section \ref{nc:theo}, we discuss the formalism used for the analytic calculations and the method for generating cross sections from the event-by-event simulation of NCQE kinematics in NEUT. We present and discuss the results of this comparison in Section \ref{results}. Finally, in Section \ref{conclusions}, we draw conclusions and provide an outlook on the future of this kind of work. 


\section{Neutral-current neutrino-nucleus scattering}
\label{nc:theo}

Let us consider the case in which a neutral current process in which a neutrino ($\nu$) or anti-neutrino ($\bar{\nu}$) scatters off a nuclear target and the hadronic final state is not detected

\begin{equation}
\nu\, (\bar{\nu}) + A \rightarrow \nu\, (\bar{\nu}) +X\ .
\end{equation}
The analytic expression of the double-differential cross section is obtained in the Born approximation as~\cite{Shen:2012xz,Benhar:2006nr}
\begin{equation}
\Big(\frac{d\sigma}{dE_{k'} d\Omega_{k'}}\Big)_{\nu/\bar{\nu}}= \frac{G_F^2 }{4\pi^2}\, k^\prime E_{k'}\, L_{\mu \nu} W^{\mu \nu}\,.
\label{eq:xsec_def}
\end{equation}
For the Fermi coupling constant we adopt the value $G_F = 1.1803 \times 10^{-5}\,\rm GeV^{-2}$, as in Ref.~\cite{Herczeg:1999}. 
With $k=(E_{k},\mathbf{k})$ and $k^\prime=(E_{k'},\mathbf{k}^\prime)$ we denote the initial and final neutrino four-momenta, respectively, and $\theta_{k'}$ 
is the neutrino scattering angle.

The leptonic tensor is fully determined by the kinematics of the leptons in the initial and final states
\begin{equation}
L_{\mu \nu}  = \frac{1}{E_{k}E_{k'}} (k_\mu k^\prime_\nu + k^\prime_\mu k_\nu - g_{\mu\nu}\, k \cdot k^\prime \pm i \epsilon_{\mu\rho\nu\sigma} k^\rho k^{\prime\, \sigma} )\, ,
\label{eq:lepton_def}
\end{equation}
where the $+$ $(-)$ sign is for $\nu$ ($\bar{\nu}$) initiated reactions. 
The hadronic tensor, describing the nuclear response to the electroweak probe, is defined in terms of the transition between the initial and final nuclear states $|\psi^A_0\rangle$ and $|\psi^A_f\rangle$, with energies $E_0$ and $E_f$ as 
\begin{align}
W^{\mu\nu}= \sum_f \langle \psi^A_0|j^{\mu \, \dagger}_{NC}(q)|\psi^A_f\rangle \langle \psi^A_f| j^\nu_{NC}(q) |\psi^A_0 \rangle \delta (E^A_0+\omega -E^A_f)\, ,
\label{eq:had_tens}
\end{align}
where $q=(\omega,{\bf q})$ is the four-momentum transferred by the probe and the neutral-current operator is the sum of a vector and axial component $j^\mu_{NC}(q) = j^\mu_{V}(q)+j^\mu_{A}(q)$.
In this work the analytic calculations presented only retain the one-body current contribution to $j^\mu_{NC}$. Its explicit expression will be discussed in Sec.~\ref{nc:operator:elem}. 

\subsection{ The Impulse Approximation and the Spectral Function Formalism}
\label{nc:ia}
At relatively large values of the momentum transfer, typically $|{\bf q}| \gtrsim 500$ MeV, the impulse approximation (IA) can be safely applied
under the assumption that the struck nucleon is decoupled from the spectator (A-1) particles~\cite{Benhar:2006wy, Benhar:2015wva}.

Employing the factorized expression of the nuclear final state
\begin{equation}
|\psi_f^A \rangle \rightarrow |p^\prime\rangle \otimes |\psi_f^{A-1}\rangle\, .
\end{equation}
and inserting a single-nucleon completeness relation, the incoherent contribution to the hadron tensor is given by
\begin{align}
W^{\mu\nu}=& \int \frac{d^3p}{(2\pi)^3} dE P_h(E,{\bf p})\frac{m_N^2}{E_p E_{p+q} }\nonumber\\
&\times  \sum_{i}\, \langle p | {j_{i,{\rm NC}}^\mu}^\dagger |p+q \rangle \langle p+q |  j_{i,{\rm NC}}^\nu | p\rangle\nonumber\\
& \times \delta(\omega+E+m_N -e(\mathbf{p+q}))\,,
\label{had:tens}
\end{align}
where $p=(E_p,{\bf p})$ is the momentum of the struck nucleon and the three-momentum conservation leads to ${\bf p}'={\bf p}+{\bf q}$ for the final nucleon.
 The factors $m_N/E_p$ and $m_N/E_{p+q}$ are included to account for the implicit covariant normalization of the four-spinors of the initial and final nucleons in the matrix elements of the relativistic current. 
The hole spectral function, $P_h(E,{\bf p})$, encompasses information on the internal nuclear structure, giving
the probability distribution of removing a nucleon with momentum ${\bf p}$ from the target nucleus, 
leaving the residual $(A-1)$-nucleon system with an excitation energy E.
The comparisons carried out in this work have been performed using two different models of hole spectral function. 

In the RFG model only statistical correlations are accounted for, leading to a very simple expression 
\begin{align}
P_h^{\text{RFG}}(E,{\bf p}\,) &= \frac{2A}{\bar\rho}\delta[E+M-(E_p-\epsilon)]\theta(\bar{p}_F-|{\bf p}\,|)\ , \nonumber \\ 
& \int \frac{d^3p}{(2\pi)^3}\,  dE\, P_h^{\text{RFG}}(E,{\bf p}\,) = 1
\label{sf_rgfg}
\end{align}
where $\bar\rho=2{p}_F^3/(3\pi^2)$ is the averaged total nucleon (both proton and neutron) density. 

In the Correlated Basis Function (CBF) approach, the presence of short-range correlated pairs is accounted for. 
The spectral function is defined as a sum of two different contributions
\begin{equation}
P^{\text{CBF}}_h(\bm{p},E) = P^{\text{MF}}(\bm{p},E) + P^{\text{Corr}}(\bm{p},E)\, .
\end{equation} 
In the first one the mean field (MF) calculation is modified by introducing spectroscopic factors and finite width functions 
extracted from (e,e'p) measurements in order to include
 the effects of residual interactions that are not present in an independent particle model description. 
The second term determines the behavior of the hole spectral function in the high momentum and removal
energy region. It has been obtained by folding CBF calculations of the spectral function in uniform and isospin symmetric
nuclear matter with the nuclear density distribution profile~\cite{Benhar:1989aw, Benhar:1994hw}.  
In the theory calculations presented in this work, one considers the initial nucleon to be bound, {\textit i.e.} the scattering process takes 
place on an off-shell nucleon. This is achieved by replacing the four momentum $q^\mu=(\omega,\vec{q}\,)$ by $ \tilde{q}^{\, \mu}=(\tilde{\omega},\vec{q}\,)$, 
such that  $\tilde{\omega}= \omega -(E_p-m_N-E)$. The modification of the energy transfer leads to a violation of the current conservation. In order to restore the gauge invariance, the De Forest prescription has been adopted~\cite{DeForest:1983ahx,Benhar:2006wy}. \\

In addition to the full analytic calculation, the RFG and CBF model spectral functions are used for the sampling of NCQE kinematics in NEUT. Details of how cross sections are generated for these models within the EG framework are given in Section \ref{EG}.

\subsection{Current operator}
\label{nc:operator:elem}
The elementary interactions for the NC processes are
\begin{align}
\label{NC:p}
\nu(k)+p(p)&\to \nu(k^\prime) + p(p^\prime)\, ,\\   
\label{NC:n}
\nu(k)+n(p)&\to \nu(k^\prime) + n(p^\prime)\, .
\end{align}
The corresponding ones for the anti-neutrino are obtained replacing 
$\nu$ with $\bar{\nu}$ both in the initial and final states. 
The one-body NC operator is the sum of a vector and axial component 
\begin{align}
&j^\mu_{NC}=j^\mu_{V}+j^\mu_{A}\nonumber\\
&j^\mu_{V}={\mathcal F}_1\gamma^\mu + i \sigma^{\mu\nu}q_\nu \frac{{\mathcal F}_2}{2m_N}\nonumber\\
&j^\mu_{A}=-\gamma^\mu \gamma_5 {\mathcal F}_A-q^\mu \gamma_5 \frac{{\mathcal F}_p}{m_N}, 
\end{align}
where
\begin{align}
{\mathcal F}_1=& \frac{1}{2}[-2\sin^2\theta_W F_1^S + (1-2\sin\theta^2_W)F_{1}^V \tau_z] \nonumber\\
{\mathcal F}_2=& \frac{1}{2}[-2\sin^2\theta_W F_2^S + (1-2\sin\theta^2_W)F_{2}^V \tau_z]\, 
\end{align}
where $\theta_W$ is the Weinberg angle ($\sin^2\theta_W$ = 0.2312~\cite{PDG}) and $\tau_z$ is the isospin operator.  
The S and V subscripts indicate the isoscalar and isovector components of the 
Dirac and Pauli form factors defined as
\begin{align}
F_{1,2}^S=&F_{1,2}^p + F_{1,2}^n\nonumber\\
F_{1,2}^V=&F_{1,2}^p - F_{1,2}^n\, . 
\end{align}
These can be expressed in terms of the Sachs form factors as
\begin{align}
F_1^{p,n}=&\frac{G_E^{p,n}+\tau G_M^{p,n}}{1+\tau}\nonumber\\
F_2^{p,n}=&\frac{G_M^{p,n}-G_E^{p,n}}{1+\tau}
\end{align}
with $\tau=-q^2/4m_N^2$. The axial term of the NC can be cast in the form
\begin{align}
{\mathcal F}_A=& F_{A} \tau_z \nonumber\\
{\mathcal F}_P=& F_{P} \tau_z\,.
\end{align}
We employ the standard dipole parametrization for the axial form factor
\begin{align}
 F_A &=\frac{g_A}{( 1- q^2/ m_A^2 )^2}\ ,
\end{align}
where the nucleon axial-vector coupling constant is taken to be $g_A=1.2694$~\cite{PDG} and the axial mass $m_A=1.03$ GeV. 
The pseudo-scalar form factor $F_P$ can be neglected when considering NC cross sections since its contribution is proportional to the mass of the final lepton. 

In order to have similar parameterizations of the neutral weak current in the analytic calculations and event generator simulations, the form of the vector and axial-vector form factors were changed from the nominal values in NEUT v5.3.6. Appendix \ref{defaults} demonstrates how each of these settings impacts the differential cross section when compared with cross sections obtained using the nominal values. The effort to have consistent neutral weak currents was important to reduce effects in the shape and normalization of the cross section arising due to changes in fundamental parameters. Hereinafter, we refer to NEUT as the alternative version of NEUT v5.3.6 with settings modified as described in Table \ref{tab:features}.

\begin{table}[h]
\begin{center}
\begin{tabular}{|c|c|} \hline
Feature &  NEUT (This Work)\\ \hline
$M_A$ (GeV)   & 1.03 \\ \hline
Vector Form Factor &  Dipole \\ \hline
$\bar{E}_B$ (MeV) &  20  \\ \hline
Pauli Blocking & Disabled \\ \hline
\end{tabular}
\label{tab:features}
\caption{Changes to features in NEUT v5.3.6 made in order to replicate the neutral weak current form factors used in the analytic theory calculations. The axial mass ($M_A$) and vector form factor of the neutral weak current, the average nucleon binding energy ($\bar{E}_B$), and the effects of Pauli blocking were changed from the default to consistently compare between the simulations and the calculations.}
\end{center}
\end{table}

\subsection{Event generator cross sections}
\label{EG}

To compare with the full calculation carried out analytically according to Section \ref{nc:ia}, simulated cross sections sampled with NEUT were created. To do so, NCQE scattering kinematics were generated on an event-by-event basis in NEUT with a Monte Carlo sampling weighted by the cross section model \cite{neut}. In the EG framework, certain kinematic variables cannot be fixed to specific values like in the factorization scheme. Instead, one must make kinematic cuts on the events to produce the desired projection. In particular, cuts were made in the scattering angle $\theta_{k'}$ to make comparisons at a fixed scattering angle. To obtain a differential cross section as a function of $\omega$, the events are first binned into histograms of bin width $\Delta\omega = 20$ MeV for NEUT. The event rates in the histogram bins can be converted into differential cross sections when scaled properly. The NUISANCE \cite{stowell2017} framework provides the weight, $W$, needed to convert the event rate into a fluxed average cross section. Because we are considering a mono-energetic beam at $E_{\nu} = 1$ GeV, the scaling factor $W$ has the form:

\begin{equation}
W=\frac{\sigma_{tot}(E_{\nu})}{N}
\label{eq:evt}
\end{equation}

Where $\sigma_{tot}(E_{\nu})$ is the total inclusive cross section at the beam energy and $N$ is the total number of events simulated. Scaling the histogram event rate centered at a given value of $\omega$, here called $N_{bin}$, by $W$ gives the flux averaged cross section per nucleon.  For a fluxed averaged cross section per target, an additional scaling by the number of nucleons in the target, $A$, is necessary. To convert this to a differential cross section, the flux-averaged cross section is divided by the bin widths in $\omega$ and the solid angle, $\Omega$. Cuts are only made on the scattering angle, resulting in a solid angle bin $\Delta\Omega = 2\pi(\cos\theta_{k'}^{min} - \cos\theta_{k'}^{max}) = 2\pi\Delta(\cos\theta_{k'})$, where $\theta_{k'}^{min/max}$ is the minimum/maximum value of the scattering angle selected for in the kinematic cuts. The double differential cross section per target for a given bin is then given by the following equation:\\

\begin{equation}
\left(\frac{d\sigma}{d\Omega d\omega}\right)_{bin} = \frac{WA}{2\pi\Delta(\cos\theta_{k'}) \times \Delta\omega} \times N_{bin} \label{eq:diffxs}
\end{equation}


\section{Results and discussion}
\label{results}

In this section, the differential cross section obtained from simulating an adequate number of NCQE events in NEUT so that the peak value of the cross sections had a sub 5\% level of uncertainty. The number of events ranged from $5\times10^{6}$ to $15\times10^{6}$ $\nu$/$\bar{\nu}$-$^{12}$ events. The cross sections generated from these simulated events were compared with analytic calculations of the same quantity using the formalism given in Section \ref{nc:ia}. Both calculations use a fixed neutrino beam energy of $E_{\text{beam}}=$ 1 GeV. 

\begin{figure}[ht]
\begin{center}
\includegraphics[width=0.45\textwidth]{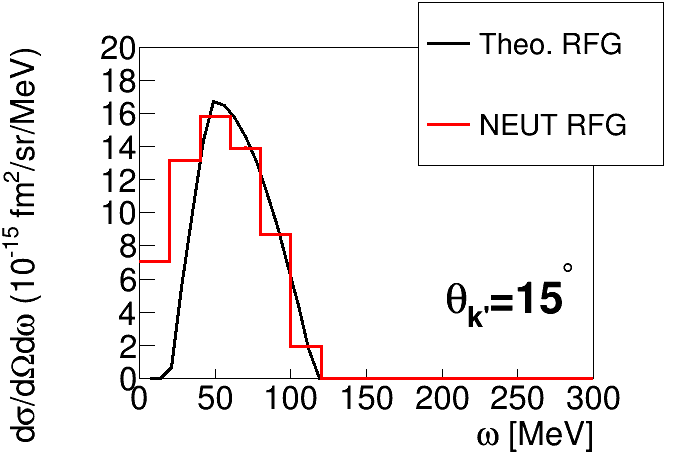}
\includegraphics[width=0.45\textwidth]{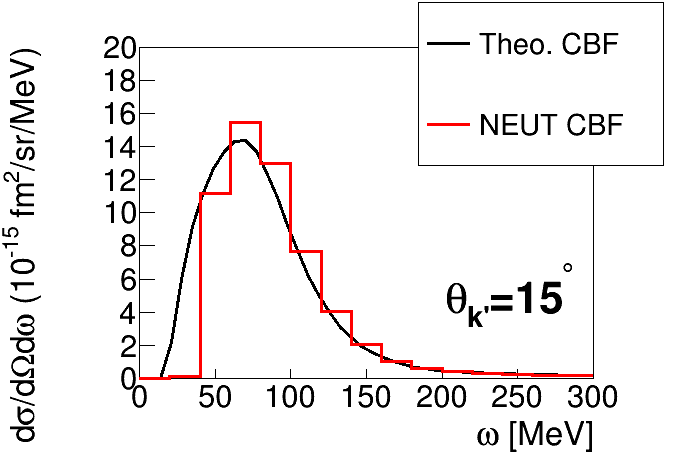}
\caption{Comparison of neutral current quasi-elastic $\nu$-$^{12}$C between NEUT (red) and the theoretical calculation (solid black line) for $\theta_{k'}= 15^{\circ}$ using the (TOP) RFG and (BOTTOM) CBF nucleon spectral functions. These comparisons are done at a fixed energy of 1 GeV. The histograms in $\omega$ are binned with widths $\Delta\omega$ = 20 MeV.}
\label{fig:nu15}
\end{center}
\end{figure}

Calculations were carried out for fixed scattering scattering angles $\theta_{k'}$ = 15$^{\circ}$, 30$^{\circ}$, 60$^{\circ}$, 70$^{\circ}$, and 120$^{\circ}$. Calculations for neutrinos using CBF and RFG at a scattering angle of $\theta_{k'}=15^{\circ}$ are shown in Figure \ref{fig:nu15} and for anti-neutrinos at the same scattering angle in Figure \ref{fig:Anu15}. Additional angles can be found in Appendix \ref{additional}.

The location of the peak of the differential cross section in energy transfer, $\omega$, is a quantity of experimental importance because of its relations to extracting the binding energy.  Because the binding energy is a parameter of the RFG model, it should be expected that the location of the peak in the cross section is reproduced when using the same value in both the EG simulation and the theoretical calculation. In most of the cases studied, the peak of the theory calculations and the peak bin of the EG simulation are consistent. Neutrino and anti-neutrino scattering at $\theta_{k'} = 30^{\circ}$ produces peak bin centers that are inconsistent with theory by 20 MeV. For the CBF model, an average binding energy parameter is not included. For the anti-neutrino induced interaction in both CBF and RFG, the peak of the theoretical calculation matches with the peak bin in the simulation results.

Tables \ref{tab:max-nu} and \ref{tab:max-Anu} report the results of the cross sections simulated with NEUT  (red) and calculated in the factorization scheme (black) using the CBF model of the differential cross sections for neutrinos and anti-neutrinos, respectively. For NCQE scattering, the differential cross section peak is consistently larger than the theoretical prediction of the same quantity using the CBF model. With the exception of $\theta_{k'}= 120^{\circ}$, the height of the differential cross section for anti-neutrino NCQE scattering in the CBF model simulated in NEUT is larger than the theory calculation. For $\theta_{k'}= 15^{\circ}, 30^{\circ}, 60^{\circ}$, and 70$^{\circ}$, the ratio between the height of the differential cross section in NEUT to the theoretical curve is consistent. In RFG, this same consistency between neutrinos and anti-neutrinos is not seen. Using RFG, NEUT over-predicts the height of the differential cross sections for all cases except $\theta_{k'}= 15^{\circ}$ for the neutrino interactions. For anti-neutrinos, the NEUT prediction of the differential cross sections using RFG is consistently lower than the theoretical ones. As the scattering angle increases, the ratio becomes increasingly smaller and the discrepancy between the calculations is significant. In particular, at $\theta_{k'}= 120^{\circ}$, the height of the NEUT differential cross section is 30\% of the theoretical value. The differences seen in the RFG distributions are potentially due to the limitation of the approximation used to calculate NCQE with the RFG model in NEUT.

\begin{figure}[ht]
\begin{center}
\includegraphics[width=0.45\textwidth]{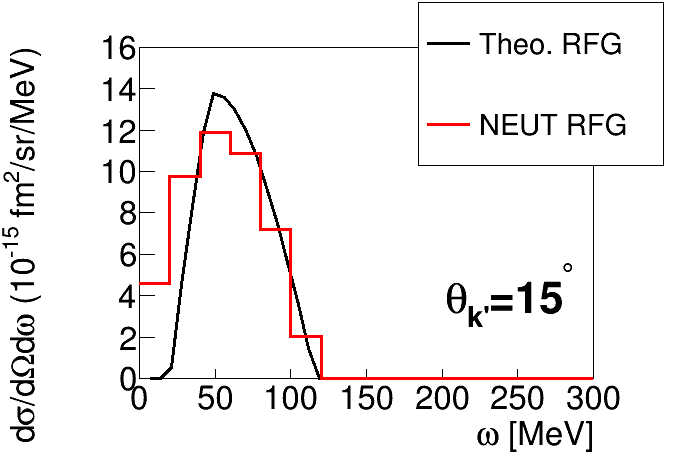}
\includegraphics[width=0.45\textwidth]{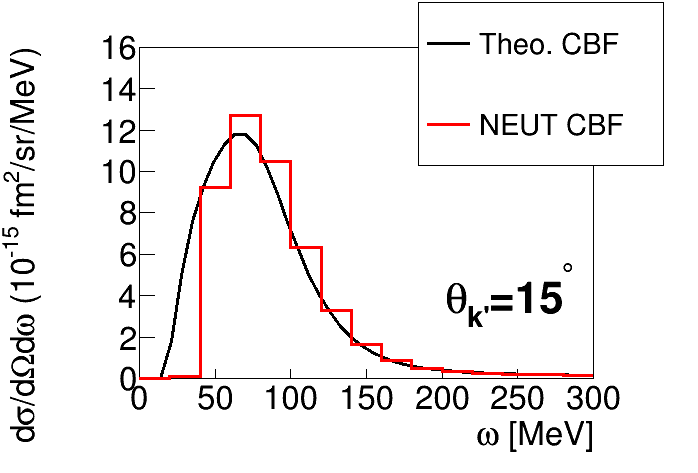}
\caption{Comparison of neutral current quasi-elastic $\bar{\nu}$-$^{12}$C between NEUT (red) and the theoretical calculation (solid black line) for $\theta_{k'}= 15^{\circ}$ using the (TOP) RFG and (BOTTOM) CBF nucleon spectral functions. These comparisons are done at a fixed energy of 1 GeV. The histograms in $\omega$ are binned with widths $\Delta\omega$ = 20 MeV.}
\label{fig:Anu15}
\end{center}
\end{figure}

\begin{table}[h]
\begin{center}
\begin{tabular}{|c|c|c|c|} \hline
Model &$\theta_{k'}$ (deg) &  $d\sigma/d\Omega d\omega_{\rm{peak}}^{\text{NEUT}}$ & $d\sigma/d\Omega d\omega_{\rm{peak}}^{\text{Theo}}$ \\ \hline  
\multirow{5}{*}{RFG} &15 & 15.81 $\pm$ 0.08 & 16.714 \\ \cline{2-4}
&30 & 5.58 $\pm$ 0.03 & 5.274 \\ \cline{2-4}
&60 & 1.07 $\pm$ 0.01 & 0.913 \\ \cline{2-4}
&70 & 0.70 $\pm$ 0.01 & 0.592 \\ \cline{2-4}
&120 & 0.193 $\pm$ 0.003 & 0.159 \\ \hline
\multirow{5}{*}{CBF} &15 & 15.46 $\pm$ 0.08 & 14.350 \\ \cline{2-4}
&30 & 5.85 $\pm$ 0.03 & 5.015 \\ \cline{2-4}
&60 & 0.10 $\pm$ 0.01 & 0.923 \\ \cline{2-4}
&70 & 0.639 $\pm$ 0.008 & 0.604 \\ \cline{2-4}
&120 & 0.173 $\pm$ 0.004 & 0.166 \\ \hline
\end{tabular}
\end{center}
\caption{Value of the peak in differential cross section using NEUT simulations and theoretical calculations of $\nu$-$^{12}$C neutral current quasi-elastic scattering at various angles. All differential cross section peak values are given in units fm$^{2}$/MeV/sr.}
\label{tab:max-nu} 
\end{table}

\begin{table}[h]
\begin{center}
\begin{tabular}{|c|c|c|c|} \hline
Model &$\theta_{k'}$ (deg) &  $d\sigma/d\Omega d\omega_{\rm{peak}}^{\text{NEUT}}$ & $d\sigma/d\Omega d\omega_{\rm{peak}}^{\text{Theo}}$ \\ \hline  
\multirow{5}{*}{RFG} &15 & 11.86 $\pm$ 0.04 & 13.782 \\ \cline{2-4}
&30 & 2.51 $\pm$ 0.01 & 2.862\\ \cline{2-4}
&60 & 0.150 $\pm$ 0.003 & 0.188 \\ \cline{2-4}
&70 & 0.066 $\pm$ 0.002 & 0.091 \\ \cline{2-4}
&120 & 0.0036 $\pm$ 0.0002 & 0.0099 \\ \hline
\multirow{5}{*}{CBF} &15 & 12.72 $\pm$ 0.05 & 11.807 \\ \cline{2-4}
&30 & 3.11 $\pm$ 0.01 & 2.704\\ \cline{2-4}
&60 & 0.210 $\pm$ 0.003 & 0.188 \\ \cline{2-4}
&70 & 0.102 $\pm$ 0.002 & 0.091 \\ \cline{2-4}
&120 & 0.0120 $\pm$ 0.0004 & 0.010 \\ \hline
\end{tabular}
\end{center}
\caption{Value of the peak in differential cross section using  NEUT simulations and theoretical calculations of $\bar{\nu}$-$^{12}$C neutral current quasi-elastic scattering at various angles. All differential cross section peak values are given in units fm$^{2}$/MeV/sr.}
\label{tab:max-Anu} 
\end{table}

An important feature of the cross section that arises when using the CBF model is the presence of a tail at large $\omega$ not seen when using only RFG. This feature arises due to interactions of the incident (anti-)neutrino with correlated nucleons and is accounted for in the correlation term in the spectral function. In order to rigorously quantify the presence of this tail, a metric of assessing the tail strength, denoted here as $r_{\rm{tail}}$, was devised. To evaluate the tail strength, the integral of the differential cross section with lower bound equal to the value where the RFG curve cross zero was calculated, called $\omega_{\rm{RFG}}^{\rm{max}}$. The percent of the total integral contained in the tail is defined as $r_{\rm{tail}}$:

\begin{equation}
r_{\rm{tail}} = 100 \% \times \int_{\omega_{\rm{RFG}}^{\rm{max}}}^{\infty} d\omega \frac{d\sigma}{d\Omega d\omega} \bigg/ \int d\omega  \frac{d\sigma}{d\Omega d\omega}
\end{equation}

Tables \ref{tab:tail-nu} and \ref{tab:tail-Anu} summarize the tail strengths of the differential cross sections for neutrinos and anti-neutrinos, respectively. In all cases, the tail strength in the EG simulations using the CBF model is less than what is seen in the theoretical calculations using this metric. While differences exist, there is some error that is introduced in finding $\omega_{\rm{RFG}}^{\rm{max}}$ due to the binning that could be responsible for this feature. The fact that the strengths of the tails are similar in magnitude is encouraging and indicates the presence of an expected behavior; however, the appearance of the tail is not the only feature one can expect to appear when using CBF. In addition to the appearance of a tail due to the correlation term in the CBF, the mean field term should be quenched. All factorization scheme calculations have a peak in the CBF calculation that is quenched or approximately equal to its RFG counterpart. For the EG simulations, this behavior is not present. For neutrino cross sections simulated with NEUT, the EG over-predicts for all angles at the peak of the CBF model compared to theory. The cross section peak at $30^{\circ}$ is enhanced rather than quenched. In all cases considered for the anti-neutrinos, all angles see an enhancement in the simulated cross sections of the CBF model relative to RFG in NEUT.

\begin{table}[h]
\begin{center}
\begin{tabular}{|c|c|c|} \hline
$\theta_{k'}$ (deg) & $r_{\rm{tail}}^{\text{NEUT}}$  (\%) & $r_{\rm{tail}}^{\text{Theo}}$  (\%) \\ \hline  
15 & 10 & 16\\ \hline
30 & 6 & 9\\ \hline
60 & 4 & 7  \\ \hline
70 & 4 & 6\\ \hline
120 & 3 & 6\\ \hline
\end{tabular}
\end{center}
\caption{Tail strength metric for the differential cross section using NEUT simulations and theoretical calculations of $\nu$-$^{12}$C neutral current quasi-elastic scattering at various angles.}
\label{tab:tail-nu} 
\end{table}

\begin{table}[h]
\begin{center}
\begin{tabular}{|c|c|c|} \hline
$\theta_{k'}$ (deg) & $r_{\rm{tail}}^{\text{NEUT}}$  (\%) & $r_{\rm{tail}}^{\text{Theo}}$  (\%) \\ \hline  
15 & 10 & 16\\ \hline
30 & 6 &  9\\ \hline
60 & 4 & 7\\ \hline
70 & 3 &  6\\ \hline
120 & 3 &  7\\ \hline
\end{tabular}
\end{center}
\caption{Tail strength metric for the differential cross section using NEUT simulations and theoretical calculations of $\bar{\nu}$-$^{12}$C neutral current quasi-elastic scattering at various angles.}
\label{tab:tail-Anu} 
\end{table}


\section{Conclusions}
\label{conclusions}

This study is the first set of comparisons between EG simulated observables of neutrino interaction cross sections with those obtained from state-of-the-art theory calculations like those presented in \cite{rocco2019}. Using the EG NEUT employed by the T2K experiment for their analysis, neutral current quasi-elastic scattering events for neutrinos and anti-neutrinos on a $^{12}$C target are simulated. 

While the EGs captured general features of the full calculation of the RFG and CBF models, it is important to note that there were still differences present between the two methods. In particular, 10-20\% differences in the cross section may lead to different predicted rates and/or modify the associated uncertainty on the interaction model relevant to oscillation and exotic physics searches. This fact demonstrates the value in benchmarking EG performance against theoretical calculations. In this way, the experimental neutrino oscillation community can hope to identify missing physics in EGs and improve simulations for the analysis of state-of-the-art measurements that will be available in the near future. In addition to neutral weak current interactions, charged current interaction observables can also be compared. Finally, comparisons to calculations using other methods of computing the spectral function and hadronic tensor should be carried out to determine the impact of using approximate models for experimental analyses.


\section{Acknowledgements}
\label{acknowledgements}
We acknowledge the support of the Office of Science, the Office of High Energy Physics, and of the U.S. Department of Energy award DE-SC0015903. 
The work of NR is supported by the U.S. Department of Energy, Office of Science, Office of Nuclear Physics, under contracts DE-AC02-06CH11357, as well as by the Nuclear Computational Low-Energy Initiative (NUCLEI) SciDAC project and by Fermi Research Alliance, LLC, under Contract No. DE-AC02-07CH11359 with the U.S. Department of Energy, Office of Science, Office of High Energy Physics. 
This work relied on computational resources provided by iCER and the High Performance Computing Center at Michigan State University. G.~B.~K. and K.~M. would like to acknowledge Alessandro Lovato, Yoshinari Hayato, Filomena Nunes, and Saori Pastore for useful discussions throughout the course of this work. G.~B.~K. would like to acknowledge support for this work from the Michigan State University College of Natural Science through the Larry D. Fowler Endowment.


\bibliography{nu}

\begin{thebibliography}{31}
\expandafter\ifx\csname natexlab\endcsname\relax\def\natexlab#1{#1}\fi
\expandafter\ifx\csname bibnamefont\endcsname\relax
  \def\bibnamefont#1{#1}\fi
\expandafter\ifx\csname bibfnamefont\endcsname\relax
  \def\bibfnamefont#1{#1}\fi
\expandafter\ifx\csname citenamefont\endcsname\relax
  \def\citenamefont#1{#1}\fi
\expandafter\ifx\csname url\endcsname\relax
  \def\url#1{\texttt{#1}}\fi
\expandafter\ifx\csname urlprefix\endcsname\relax\def\urlprefix{URL }\fi
\providecommand{\bibinfo}[2]{#2}
\providecommand{\eprint}[2][]{\url{#2}}

\bibitem[{\citenamefont{Rocco et~al.}(2019)\citenamefont{Rocco, Barbieri,
  Benhar, De~Pace, and Lovato}}]{rocco2019}
\bibinfo{author}{\bibfnamefont{N.}~\bibnamefont{Rocco}},
  \bibinfo{author}{\bibfnamefont{C.}~\bibnamefont{Barbieri}},
  \bibinfo{author}{\bibfnamefont{O.}~\bibnamefont{Benhar}},
  \bibinfo{author}{\bibfnamefont{A.}~\bibnamefont{De~Pace}}, \bibnamefont{and}
  \bibinfo{author}{\bibfnamefont{A.}~\bibnamefont{Lovato}},
  \bibinfo{journal}{Phys. Rev. C} \textbf{\bibinfo{volume}{99}},
  \bibinfo{pages}{025502} (\bibinfo{year}{2019}),
  \urlprefix\url{https://link.aps.org/doi/10.1103/PhysRevC.99.025502}.

\bibitem[{\citenamefont{Benhar et~al.}(1994{\natexlab{a}})\citenamefont{Benhar,
  Fabrocini, Fantoni, and Sick}}]{benhar1994}
\bibinfo{author}{\bibfnamefont{O.}~\bibnamefont{Benhar}},
  \bibinfo{author}{\bibfnamefont{A.}~\bibnamefont{Fabrocini}},
  \bibinfo{author}{\bibfnamefont{S.}~\bibnamefont{Fantoni}}, \bibnamefont{and}
  \bibinfo{author}{\bibfnamefont{I.}~\bibnamefont{Sick}},
  \bibinfo{journal}{Nuclear Physics A} \textbf{\bibinfo{volume}{579}},
  \bibinfo{pages}{493 } (\bibinfo{year}{1994}{\natexlab{a}}), ISSN
  \bibinfo{issn}{0375-9474},
  \urlprefix\url{http://www.sciencedirect.com/science/article/pii/0375947494909202}.

\bibitem[{t2k()}]{t2k}
\emph{\bibinfo{title}{{The T2K Experiment}}},
  \urlprefix\url{http://t2k-experiment.org/}.

\bibitem[{nov()}]{nova}
\emph{\bibinfo{title}{{The NO$\nu$A Experiment}}},
  \urlprefix\url{http://novaexperiment.fnal.gov/}.

\bibitem[{mic()}]{microboone}
\emph{\bibinfo{title}{{The MicroBooNE Experiment}}},
  \urlprefix\url{http://microboone.fnal.gov/}.

\bibitem[{min()}]{minerva}
\emph{\bibinfo{title}{{The MINER$\nu$A Experiment}}},
  \urlprefix\url{http://minerva.fnal.gov/}.

\bibitem[{dun()}]{dune}
\emph{\bibinfo{title}{{The Deep Underground Neutrino Experiment}}},
  \urlprefix\url{http://www.dunescience.org/}.

\bibitem[{hyp()}]{hyperk}
\emph{\bibinfo{title}{{Hyper-Kamiokande}}},
  \urlprefix\url{http://www.hyperk.org/}.

\bibitem[{\citenamefont{Alvarez-Ruso et~al.}(2018)\citenamefont{Alvarez-Ruso,
  Athar, Barbaro, Cherdack, Christy, Coloma, Donnelly, Dytman, de~Gouvêa, Hill
  et~al.}}]{nustec}
\bibinfo{author}{\bibfnamefont{L.}~\bibnamefont{Alvarez-Ruso}},
  \bibinfo{author}{\bibfnamefont{M.~S.} \bibnamefont{Athar}},
  \bibinfo{author}{\bibfnamefont{M.}~\bibnamefont{Barbaro}},
  \bibinfo{author}{\bibfnamefont{D.}~\bibnamefont{Cherdack}},
  \bibinfo{author}{\bibfnamefont{M.}~\bibnamefont{Christy}},
  \bibinfo{author}{\bibfnamefont{P.}~\bibnamefont{Coloma}},
  \bibinfo{author}{\bibfnamefont{T.}~\bibnamefont{Donnelly}},
  \bibinfo{author}{\bibfnamefont{S.}~\bibnamefont{Dytman}},
  \bibinfo{author}{\bibfnamefont{A.}~\bibnamefont{de~Gouvêa}},
  \bibinfo{author}{\bibfnamefont{R.}~\bibnamefont{Hill}}, \bibnamefont{et~al.},
  \bibinfo{journal}{Progress in Particle and Nuclear Physics}
  \textbf{\bibinfo{volume}{100}}, \bibinfo{pages}{1 } (\bibinfo{year}{2018}),
  ISSN \bibinfo{issn}{0146-6410},
  \urlprefix\url{http://www.sciencedirect.com/science/article/pii/S0146641018300061}.

\bibitem[{\citenamefont{Adamson et~al.}(2011)\citenamefont{Adamson, Auty,
  Ayres, Backhouse, Barr, Bishai, Blake, Bock, Boehnlein, Bogert
  et~al.}}]{minos2011}
\bibinfo{author}{\bibfnamefont{P.}~\bibnamefont{Adamson}},
  \bibinfo{author}{\bibfnamefont{D.~J.} \bibnamefont{Auty}},
  \bibinfo{author}{\bibfnamefont{D.~S.} \bibnamefont{Ayres}},
  \bibinfo{author}{\bibfnamefont{C.}~\bibnamefont{Backhouse}},
  \bibinfo{author}{\bibfnamefont{G.}~\bibnamefont{Barr}},
  \bibinfo{author}{\bibfnamefont{M.}~\bibnamefont{Bishai}},
  \bibinfo{author}{\bibfnamefont{A.}~\bibnamefont{Blake}},
  \bibinfo{author}{\bibfnamefont{G.~J.} \bibnamefont{Bock}},
  \bibinfo{author}{\bibfnamefont{D.~J.} \bibnamefont{Boehnlein}},
  \bibinfo{author}{\bibfnamefont{D.}~\bibnamefont{Bogert}},
  \bibnamefont{et~al.} (\bibinfo{collaboration}{MINOS Collaboration}),
  \bibinfo{journal}{Phys. Rev. Lett.} \textbf{\bibinfo{volume}{107}},
  \bibinfo{pages}{011802} (\bibinfo{year}{2011}),
  \urlprefix\url{https://link.aps.org/doi/10.1103/PhysRevLett.107.011802}.

\bibitem[{\citenamefont{Adamson et~al.}(2017)\citenamefont{Adamson, Aliaga,
  Ambrose, Anfimov, Antoshkin, Arrieta-Diaz, Augsten, Aurisano, Backhouse,
  Baird et~al.}}]{nova2017}
\bibinfo{author}{\bibfnamefont{P.}~\bibnamefont{Adamson}},
  \bibinfo{author}{\bibfnamefont{L.}~\bibnamefont{Aliaga}},
  \bibinfo{author}{\bibfnamefont{D.}~\bibnamefont{Ambrose}},
  \bibinfo{author}{\bibfnamefont{N.}~\bibnamefont{Anfimov}},
  \bibinfo{author}{\bibfnamefont{A.}~\bibnamefont{Antoshkin}},
  \bibinfo{author}{\bibfnamefont{E.}~\bibnamefont{Arrieta-Diaz}},
  \bibinfo{author}{\bibfnamefont{K.}~\bibnamefont{Augsten}},
  \bibinfo{author}{\bibfnamefont{A.}~\bibnamefont{Aurisano}},
  \bibinfo{author}{\bibfnamefont{C.}~\bibnamefont{Backhouse}},
  \bibinfo{author}{\bibfnamefont{M.}~\bibnamefont{Baird}}, \bibnamefont{et~al.}
  (\bibinfo{collaboration}{The NOvA Collaboration}), \bibinfo{journal}{Phys.
  Rev. D} \textbf{\bibinfo{volume}{96}}, \bibinfo{pages}{072006}
  (\bibinfo{year}{2017}),
  \urlprefix\url{https://link.aps.org/doi/10.1103/PhysRevD.96.072006}.

\bibitem[{\citenamefont{Abe et~al.}(2019)\citenamefont{Abe, Akutsu, Ali,
  Andreopoulos, Anthony, Antonova, Aoki, Ariga, Ashida, Awataguchi
  et~al.}}]{t2k2019}
\bibinfo{author}{\bibfnamefont{K.}~\bibnamefont{Abe}},
  \bibinfo{author}{\bibfnamefont{R.}~\bibnamefont{Akutsu}},
  \bibinfo{author}{\bibfnamefont{A.}~\bibnamefont{Ali}},
  \bibinfo{author}{\bibfnamefont{C.}~\bibnamefont{Andreopoulos}},
  \bibinfo{author}{\bibfnamefont{L.}~\bibnamefont{Anthony}},
  \bibinfo{author}{\bibfnamefont{M.}~\bibnamefont{Antonova}},
  \bibinfo{author}{\bibfnamefont{S.}~\bibnamefont{Aoki}},
  \bibinfo{author}{\bibfnamefont{A.}~\bibnamefont{Ariga}},
  \bibinfo{author}{\bibfnamefont{Y.}~\bibnamefont{Ashida}},
  \bibinfo{author}{\bibfnamefont{Y.}~\bibnamefont{Awataguchi}},
  \bibnamefont{et~al.} (\bibinfo{collaboration}{T2K Collaboration}),
  \bibinfo{journal}{Phys. Rev. D} \textbf{\bibinfo{volume}{99}},
  \bibinfo{pages}{071103} (\bibinfo{year}{2019}),
  \urlprefix\url{https://link.aps.org/doi/10.1103/PhysRevD.99.071103}.

\bibitem[{\citenamefont{Hayato}(2009)}]{neut}
\bibinfo{author}{\bibfnamefont{Y.}~\bibnamefont{Hayato}},
  \bibinfo{journal}{Acta Phys. Polon.} \textbf{\bibinfo{volume}{B40}},
  \bibinfo{pages}{2477} (\bibinfo{year}{2009}).

\bibitem[{\citenamefont{Furmanski}(2015)}]{furmanski2015}
\bibinfo{author}{\bibfnamefont{A.~P.} \bibnamefont{Furmanski}}, Ph.D. thesis,
  \bibinfo{school}{University of Warwick} (\bibinfo{year}{2015}).

\bibitem[{\citenamefont{Lovato et~al.}(2018)\citenamefont{Lovato, Gandolfi,
  Carlson, Lusk, Pieper, and Schiavilla}}]{lovato2018}
\bibinfo{author}{\bibfnamefont{A.}~\bibnamefont{Lovato}},
  \bibinfo{author}{\bibfnamefont{S.}~\bibnamefont{Gandolfi}},
  \bibinfo{author}{\bibfnamefont{J.}~\bibnamefont{Carlson}},
  \bibinfo{author}{\bibfnamefont{E.}~\bibnamefont{Lusk}},
  \bibinfo{author}{\bibfnamefont{S.~C.} \bibnamefont{Pieper}},
  \bibnamefont{and}
  \bibinfo{author}{\bibfnamefont{R.}~\bibnamefont{Schiavilla}},
  \bibinfo{journal}{Phys. Rev. C} \textbf{\bibinfo{volume}{97}},
  \bibinfo{pages}{022502} (\bibinfo{year}{2018}),
  \urlprefix\url{https://link.aps.org/doi/10.1103/PhysRevC.97.022502}.

\bibitem[{\citenamefont{Carlson et~al.}(2015)\citenamefont{Carlson, Gandolfi,
  Pederiva, Pieper, Schiavilla, Schmidt, and Wiringa}}]{carlson2015}
\bibinfo{author}{\bibfnamefont{J.}~\bibnamefont{Carlson}},
  \bibinfo{author}{\bibfnamefont{S.}~\bibnamefont{Gandolfi}},
  \bibinfo{author}{\bibfnamefont{F.}~\bibnamefont{Pederiva}},
  \bibinfo{author}{\bibfnamefont{S.~C.} \bibnamefont{Pieper}},
  \bibinfo{author}{\bibfnamefont{R.}~\bibnamefont{Schiavilla}},
  \bibinfo{author}{\bibfnamefont{K.~E.} \bibnamefont{Schmidt}},
  \bibnamefont{and} \bibinfo{author}{\bibfnamefont{R.~B.}
  \bibnamefont{Wiringa}}, \bibinfo{journal}{Rev. Mod. Phys.}
  \textbf{\bibinfo{volume}{87}}, \bibinfo{pages}{1067} (\bibinfo{year}{2015}),
  \urlprefix\url{https://link.aps.org/doi/10.1103/RevModPhys.87.1067}.

\bibitem[{\citenamefont{Wiringa et~al.}(1995)\citenamefont{Wiringa, Stoks, and
  Schiavilla}}]{wiringa1995}
\bibinfo{author}{\bibfnamefont{R.~B.} \bibnamefont{Wiringa}},
  \bibinfo{author}{\bibfnamefont{V.~G.~J.} \bibnamefont{Stoks}},
  \bibnamefont{and}
  \bibinfo{author}{\bibfnamefont{R.}~\bibnamefont{Schiavilla}},
  \bibinfo{journal}{Phys. Rev. C} \textbf{\bibinfo{volume}{51}},
  \bibinfo{pages}{38} (\bibinfo{year}{1995}),
  \urlprefix\url{https://link.aps.org/doi/10.1103/PhysRevC.51.38}.

\bibitem[{\citenamefont{Pieper and Wiringa}(2001)}]{pieper2001}
\bibinfo{author}{\bibfnamefont{S.~C.} \bibnamefont{Pieper}} \bibnamefont{and}
  \bibinfo{author}{\bibfnamefont{R.~B.} \bibnamefont{Wiringa}},
  \bibinfo{journal}{Annual Review of Nuclear and Particle Science}
  \textbf{\bibinfo{volume}{51}}, \bibinfo{pages}{53} (\bibinfo{year}{2001}),
  \eprint{https://doi.org/10.1146/annurev.nucl.51.101701.132506},
  \urlprefix\url{https://doi.org/10.1146/annurev.nucl.51.101701.132506}.

\bibitem[{\citenamefont{Dickhoff and Barbieri}(2004)}]{dickhoff2004}
\bibinfo{author}{\bibfnamefont{W.}~\bibnamefont{Dickhoff}} \bibnamefont{and}
  \bibinfo{author}{\bibfnamefont{C.}~\bibnamefont{Barbieri}},
  \bibinfo{journal}{Progress in Particle and Nuclear Physics}
  \textbf{\bibinfo{volume}{52}}, \bibinfo{pages}{377 } (\bibinfo{year}{2004}),
  ISSN \bibinfo{issn}{0146-6410},
  \urlprefix\url{http://www.sciencedirect.com/science/article/pii/S0146641004000535}.

\bibitem[{\citenamefont{Barbieri}(2014)}]{barbieri2014}
\bibinfo{author}{\bibfnamefont{C.}~\bibnamefont{Barbieri}},
  \bibinfo{journal}{Journal of Physics: Conference Series}
  \textbf{\bibinfo{volume}{529}}, \bibinfo{pages}{012005}
  (\bibinfo{year}{2014}),
  \urlprefix\url{https://doi.org/10.1088%2F1742-6596%2F529%2F1%2F012005}.

\bibitem[{\citenamefont{Shen et~al.}(2012)\citenamefont{Shen, Marcucci,
  Carlson, Gandolfi, and Schiavilla}}]{Shen:2012xz}
\bibinfo{author}{\bibfnamefont{G.}~\bibnamefont{Shen}},
  \bibinfo{author}{\bibfnamefont{L.~E.} \bibnamefont{Marcucci}},
  \bibinfo{author}{\bibfnamefont{J.}~\bibnamefont{Carlson}},
  \bibinfo{author}{\bibfnamefont{S.}~\bibnamefont{Gandolfi}}, \bibnamefont{and}
  \bibinfo{author}{\bibfnamefont{R.}~\bibnamefont{Schiavilla}},
  \bibinfo{journal}{Phys. Rev.} \textbf{\bibinfo{volume}{C86}},
  \bibinfo{pages}{035503} (\bibinfo{year}{2012}), \eprint{1205.4337}.

\bibitem[{\citenamefont{Benhar and Meloni}(2007)}]{Benhar:2006nr}
\bibinfo{author}{\bibfnamefont{O.}~\bibnamefont{Benhar}} \bibnamefont{and}
  \bibinfo{author}{\bibfnamefont{D.}~\bibnamefont{Meloni}},
  \bibinfo{journal}{Nucl. Phys.} \textbf{\bibinfo{volume}{A789}},
  \bibinfo{pages}{379} (\bibinfo{year}{2007}), \eprint{hep-ph/0610403}.

\bibitem[{\citenamefont{Herczeg et~al.}(1999)\citenamefont{Herczeg, Hoffman,
  and Klapdor-Kleingrothaus}}]{Herczeg:1999}
\bibinfo{author}{\bibfnamefont{P.}~\bibnamefont{Herczeg}},
  \bibinfo{author}{\bibfnamefont{C.~M.} \bibnamefont{Hoffman}},
  \bibnamefont{and} \bibinfo{author}{\bibfnamefont{H.~V.}
  \bibnamefont{Klapdor-Kleingrothaus}}, \emph{\bibinfo{title}{Physics Beyond
  the Standard Model}} (\bibinfo{year}{1999}), pp. \bibinfo{pages}{1--802},
  \urlprefix\url{https://www.worldscientific.com/doi/abs/10.1142/9789814527514}.

\bibitem[{\citenamefont{Benhar et~al.}(2008)\citenamefont{Benhar, Day, and
  Sick}}]{Benhar:2006wy}
\bibinfo{author}{\bibfnamefont{O.}~\bibnamefont{Benhar}},
  \bibinfo{author}{\bibfnamefont{D.}~\bibnamefont{Day}}, \bibnamefont{and}
  \bibinfo{author}{\bibfnamefont{I.}~\bibnamefont{Sick}},
  \bibinfo{journal}{Rev. Mod. Phys.} \textbf{\bibinfo{volume}{80}},
  \bibinfo{pages}{189} (\bibinfo{year}{2008}), \eprint{nucl-ex/0603029}.

\bibitem[{\citenamefont{Benhar et~al.}(2017)\citenamefont{Benhar, Huber,
  Mariani, and Meloni}}]{Benhar:2015wva}
\bibinfo{author}{\bibfnamefont{O.}~\bibnamefont{Benhar}},
  \bibinfo{author}{\bibfnamefont{P.}~\bibnamefont{Huber}},
  \bibinfo{author}{\bibfnamefont{C.}~\bibnamefont{Mariani}}, \bibnamefont{and}
  \bibinfo{author}{\bibfnamefont{D.}~\bibnamefont{Meloni}},
  \bibinfo{journal}{Phys. Rept.} \textbf{\bibinfo{volume}{700}},
  \bibinfo{pages}{1} (\bibinfo{year}{2017}), \eprint{1501.06448}.

\bibitem[{\citenamefont{Benhar et~al.}(1989)\citenamefont{Benhar, Fabrocini,
  and Fantoni}}]{Benhar:1989aw}
\bibinfo{author}{\bibfnamefont{O.}~\bibnamefont{Benhar}},
  \bibinfo{author}{\bibfnamefont{A.}~\bibnamefont{Fabrocini}},
  \bibnamefont{and} \bibinfo{author}{\bibfnamefont{S.}~\bibnamefont{Fantoni}},
  \bibinfo{journal}{Nucl. Phys.} \textbf{\bibinfo{volume}{A505}},
  \bibinfo{pages}{267} (\bibinfo{year}{1989}).

\bibitem[{\citenamefont{Benhar et~al.}(1994{\natexlab{b}})\citenamefont{Benhar,
  Fabrocini, Fantoni, and Sick}}]{Benhar:1994hw}
\bibinfo{author}{\bibfnamefont{O.}~\bibnamefont{Benhar}},
  \bibinfo{author}{\bibfnamefont{A.}~\bibnamefont{Fabrocini}},
  \bibinfo{author}{\bibfnamefont{S.}~\bibnamefont{Fantoni}}, \bibnamefont{and}
  \bibinfo{author}{\bibfnamefont{I.}~\bibnamefont{Sick}},
  \bibinfo{journal}{Nucl.\ Phys.\ A} \textbf{\bibinfo{volume}{579}},
  \bibinfo{pages}{493} (\bibinfo{year}{1994}{\natexlab{b}}).

\bibitem[{\citenamefont{De~Forest}(1983)}]{DeForest:1983ahx}
\bibinfo{author}{\bibfnamefont{T.}~\bibnamefont{De~Forest}},
  \bibinfo{journal}{Nucl. Phys.} \textbf{\bibinfo{volume}{A392}},
  \bibinfo{pages}{232} (\bibinfo{year}{1983}).

\bibitem[{\citenamefont{Nakamura and Group}(2010)}]{PDG}
\bibinfo{author}{\bibfnamefont{K.}~\bibnamefont{Nakamura}} \bibnamefont{and}
  \bibinfo{author}{\bibfnamefont{P.~D.} \bibnamefont{Group}},
  \bibinfo{journal}{Journal of Physics G: Nuclear and Particle Physics}
  \textbf{\bibinfo{volume}{37}}, \bibinfo{pages}{075021}
  (\bibinfo{year}{2010}).

\bibitem[{\citenamefont{Bradford et~al.}(2006)\citenamefont{Bradford, Bodek,
  Budd, and Arrington}}]{bradford2006}
\bibinfo{author}{\bibfnamefont{R.}~\bibnamefont{Bradford}},
  \bibinfo{author}{\bibfnamefont{A.}~\bibnamefont{Bodek}},
  \bibinfo{author}{\bibfnamefont{H.}~\bibnamefont{Budd}}, \bibnamefont{and}
  \bibinfo{author}{\bibfnamefont{J.}~\bibnamefont{Arrington}},
  \bibinfo{journal}{Nuclear Physics B - Proceedings Supplements}
  \textbf{\bibinfo{volume}{159}}, \bibinfo{pages}{127 } (\bibinfo{year}{2006}),
  ISSN \bibinfo{issn}{0920-5632}, \bibinfo{note}{proceedings of the 4th
  International Workshop on Neutrino-Nucleus Interactions in the Few-GeV
  Region},
  \urlprefix\url{http://www.sciencedirect.com/science/article/pii/S0920563206005184}.

\bibitem[{\citenamefont{Stowell et~al.}(2017)\citenamefont{Stowell, Wret,
  Wilkinson, Pickering, Cartwright, Hayato, Mahn, McFarland, Sobczyk, Terri
  et~al.}}]{stowell2017}
\bibinfo{author}{\bibfnamefont{P.}~\bibnamefont{Stowell}},
  \bibinfo{author}{\bibfnamefont{C.}~\bibnamefont{Wret}},
  \bibinfo{author}{\bibfnamefont{C.}~\bibnamefont{Wilkinson}},
  \bibinfo{author}{\bibfnamefont{L.}~\bibnamefont{Pickering}},
  \bibinfo{author}{\bibfnamefont{S.}~\bibnamefont{Cartwright}},
  \bibinfo{author}{\bibfnamefont{Y.}~\bibnamefont{Hayato}},
  \bibinfo{author}{\bibfnamefont{K.}~\bibnamefont{Mahn}},
  \bibinfo{author}{\bibfnamefont{K.}~\bibnamefont{McFarland}},
  \bibinfo{author}{\bibfnamefont{J.}~\bibnamefont{Sobczyk}},
  \bibinfo{author}{\bibfnamefont{R.}~\bibnamefont{Terri}},
  \bibnamefont{et~al.}, \bibinfo{journal}{Journal of Instrumentation}
  \textbf{\bibinfo{volume}{12}}, \bibinfo{pages}{P01016}
  (\bibinfo{year}{2017}),
  \urlprefix\url{http://stacks.iop.org/1748-0221/12/i=01/a=P01016}.

\end{thebibliography}

\begin{appendix}
\section{Changing NEUT defaults}
\label{defaults}

\begin{figure*}[ht]
\begin{center}
\includegraphics[width=0.3\textwidth]{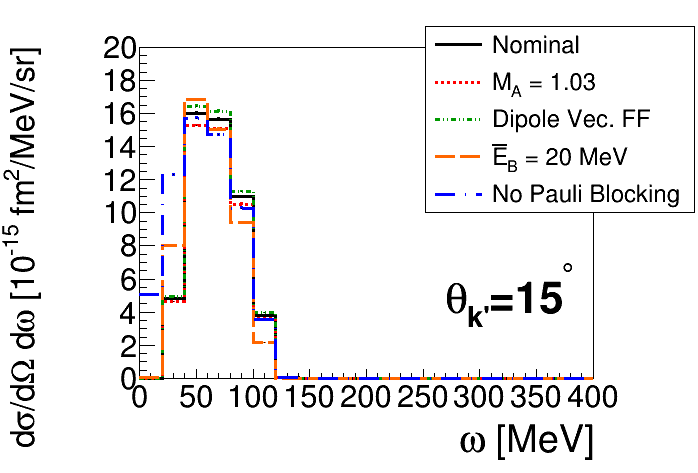}
\includegraphics[width=0.3\textwidth]{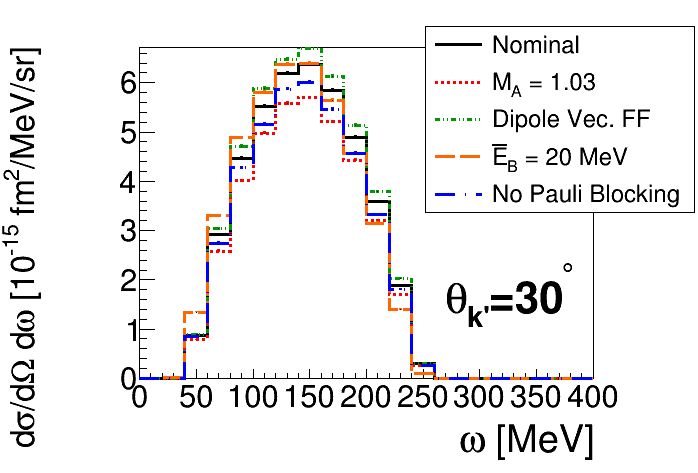}\\
\includegraphics[width=0.3\textwidth]{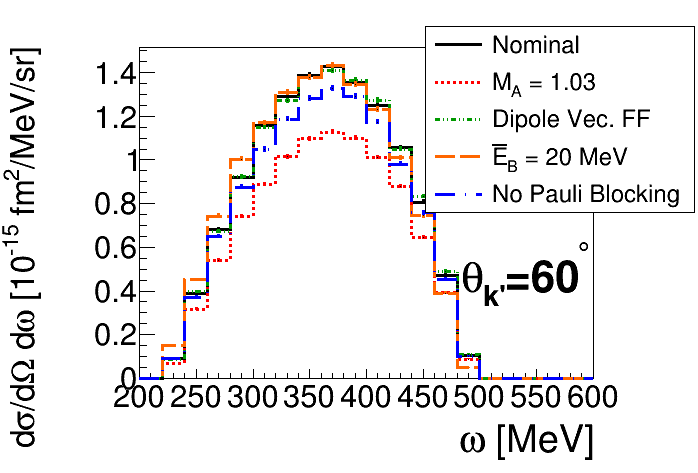}
\includegraphics[width=0.3\textwidth]{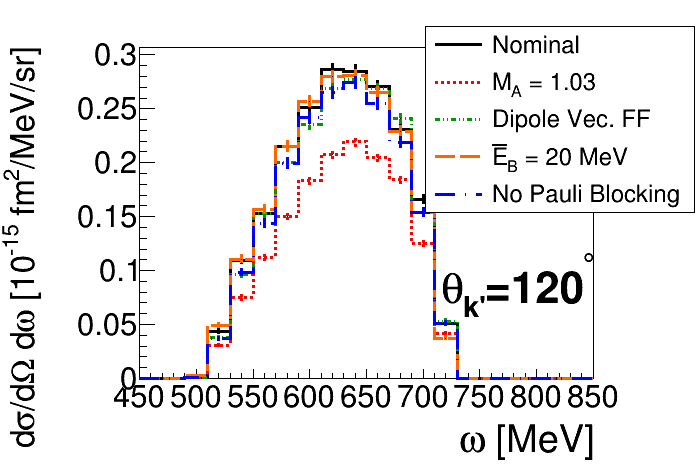}
\end{center}
\caption{Comparison of neutral current quasi-elastic $\nu$-$^{12}$C with various changes to the NEUT v5.3.6 default parameters. The black line is the nominal value in NEUT. The red line uses $M_A$ = 1.03; the green line has a dipole vector form factor; the orange line uses a nucleon binding energy $\bar{E}_B$ = 20 MeV; the blue line has Pauli blocking disabled. These comparisons are done at a fixed energy of 1 GeV using the RFG model for nucleon dynamics. The histograms in $\omega$ are binned with widths $\Delta\omega$ = 20 MeV}
\label{fig:comp-nu-rf}
\end{figure*}

\begin{figure*}[ht]
\begin{center}
\includegraphics[width=0.3\textwidth]{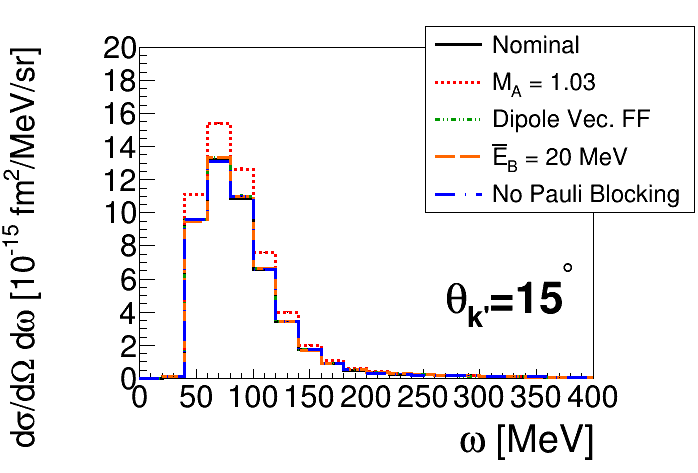}
\includegraphics[width=0.3\textwidth]{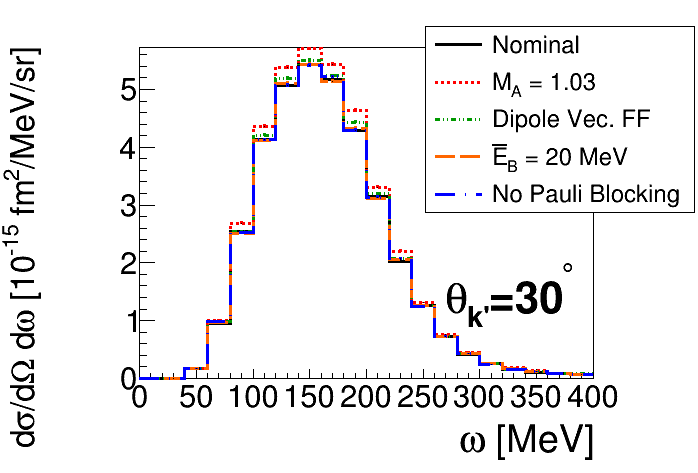}\\
\includegraphics[width=0.3\textwidth]{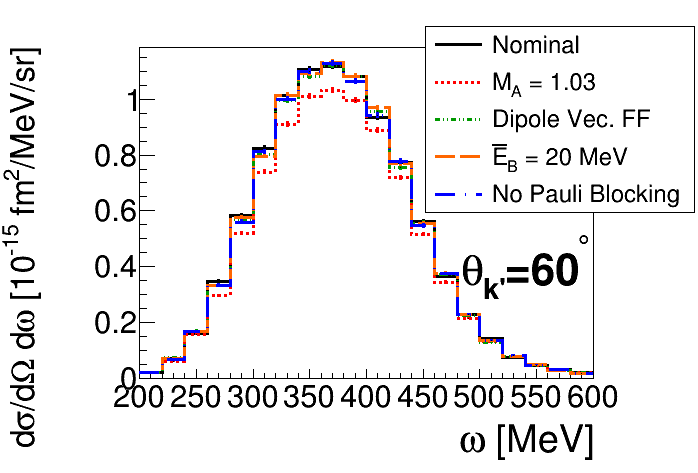}
\includegraphics[width=0.3\textwidth]{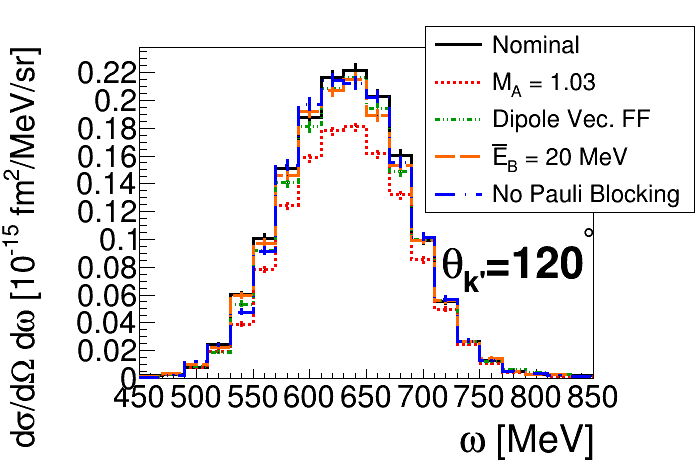}
\end{center}
\caption{Comparison of neutral current quasi-elastic $\nu$-$^{12}$C with various changes to the NEUT v5.3.6 default parameters. The black line is the nominal value in NEUT. The red line uses $M_A$ = 1.03; the green line has a dipole vector form factor; the orange line uses a nucleon binding energy $\bar{E}_B$ = 20 MeV; the blue line has Pauli blocking disabled. These comparisons are done at a fixed energy of 1 GeV using the CBF model for nucleon dynamics. The histograms in $\omega$ are binned with widths $\Delta\omega$ = 20 MeV}
\label{fig:comp-nu-sf}
\end{figure*}

\begin{figure*}[ht]
\begin{center}
\includegraphics[width=0.3\textwidth]{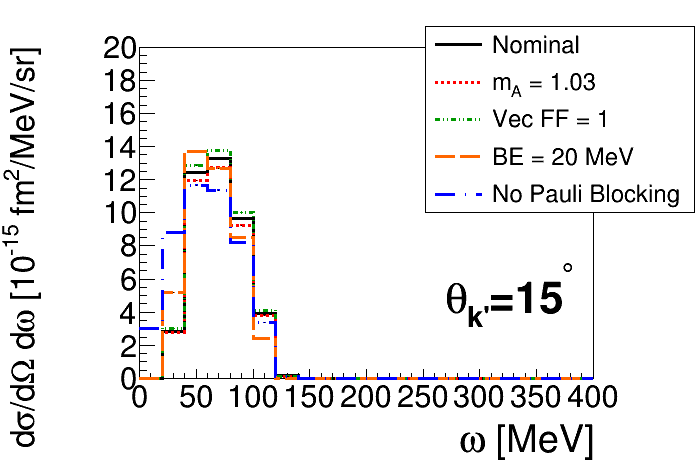}
\includegraphics[width=0.3\textwidth]{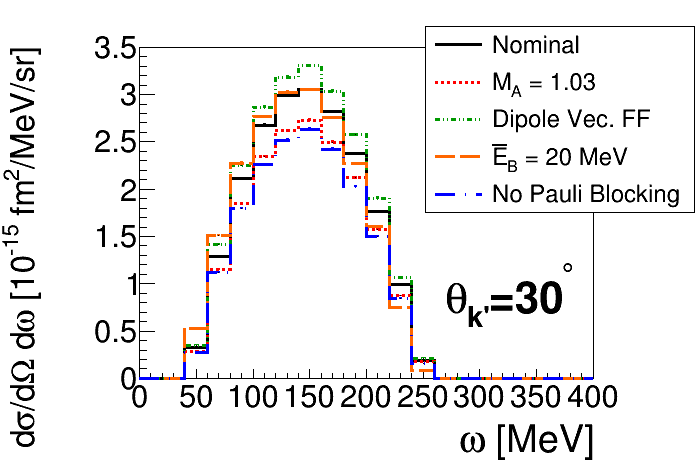}\\
\includegraphics[width=0.3\textwidth]{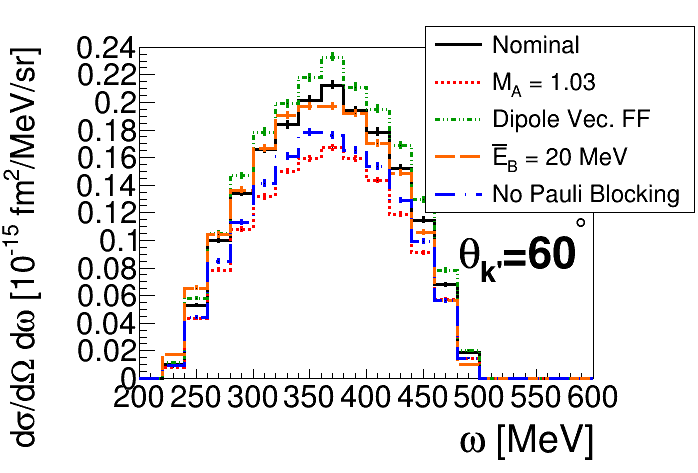}
\includegraphics[width=0.3\textwidth]{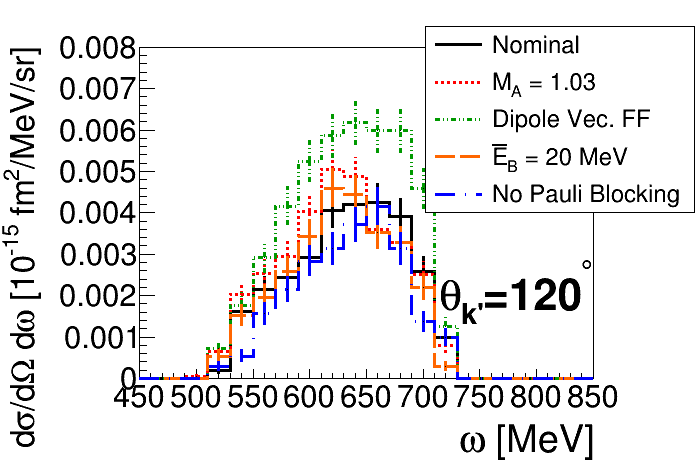}
\end{center}
\caption{Comparison of neutral current quasi-elastic $\bar{\nu}$-$^{12}$C with various changes to the NEUT v5.3.6 default parameters. The black line is the nominal value in NEUT. The red line uses $M_A$ = 1.03; the green line has a dipole vector form factor; the orange line uses a nucleon binding energy $\bar{E}_B$ = 20 MeV; the blue line has Pauli blocking disabled. These comparisons are done at a fixed energy of 1 GeV using the RFG model for nucleon dynamics. The histograms in $\omega$ are binned with widths $\Delta\omega$ = 20 MeV}
\label{fig:comp-Anu-rf}
\end{figure*}

\begin{figure*}[ht]
\begin{center}
\includegraphics[width=0.3\textwidth]{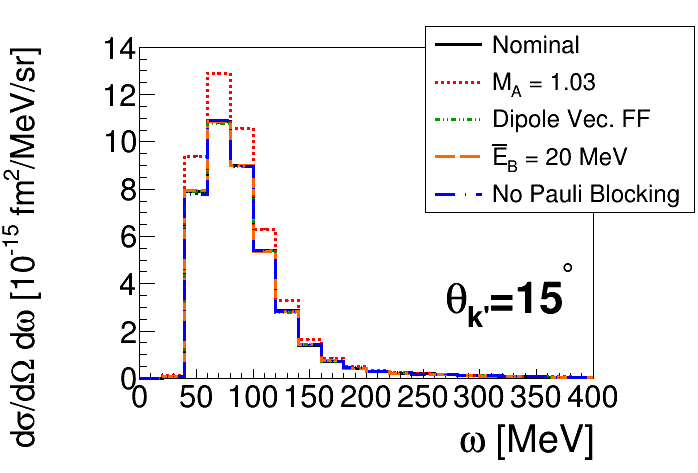}
\includegraphics[width=0.3\textwidth]{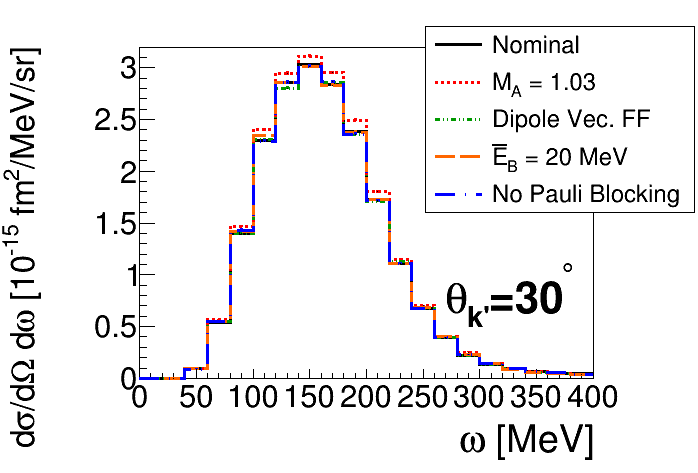}\\
\includegraphics[width=0.3\textwidth]{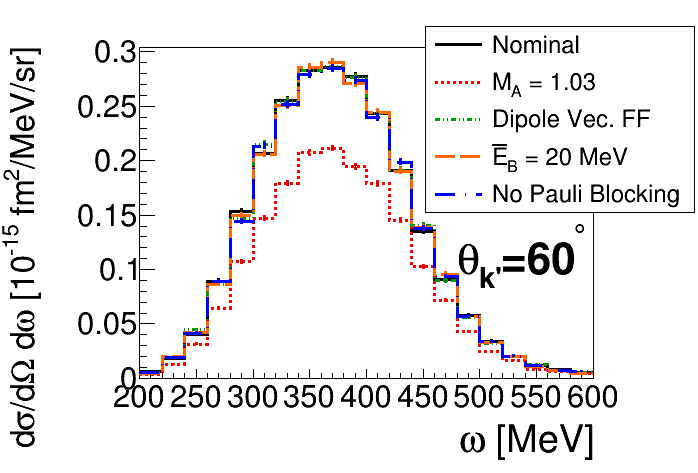}
\includegraphics[width=0.3\textwidth]{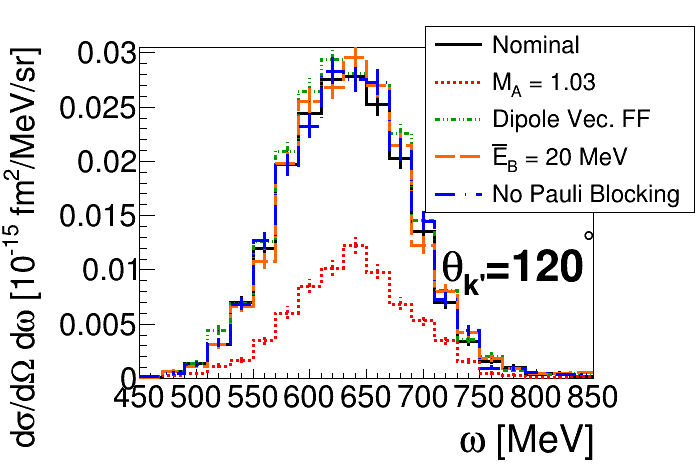}
\end{center}
\caption{Comparison of neutral current quasi-elastic $\bar{\nu}$-$^{12}$C with various changes to the NEUT v5.3.6 default parameters. The black line is the nominal value in NEUT. The red line uses $M_A$ = 1.03; the green line has a dipole vector form factor; the orange line uses a nucleon binding energy $\bar{E}_B$ = 20 MeV; the blue line has Pauli blocking disabled. These comparisons are done at a fixed energy of 1 GeV using the CBF model for nucleon dynamics. The histograms in $\omega$ are binned with widths $\Delta\omega$ = 20 MeV}
\label{fig:comp-Anu-sf}
\end{figure*}

For the comparisons with the theoretical calculations, changes were made to the nominal features of NEUT v5.3.6. The figures analyzed in this work show curves with four features of NEUT v5.3.6 changed simultaneously. Here the impact of each parameter is investigated by altering one parameter at a time and analyzing how it changes the cross sections simulated with NEUT. This is done both to demonstrate the motivation of changing parameters and to check that certain physical expectations are met by the event generator. Explorations of this sort were used as a validation before making comparisons between NEUT simulations and theoretical calculations. Figures \ref{fig:comp-nu-rf} to \ref{fig:comp-Anu-sf} show the curves analyzed in this section.

The axial mass, a parameter of the neutral current, was changed in NEUT to be consistent with that used in the theoretical curves. For neutrinos, lowering this parameter from the NEUT v3.5.6 default results in a change in the normalization of the curve. Overall, the curve is consistently lower for all angles using a lower value of the axial mass. In the case of anti-neutrinos, this same lowering is seen for $\theta_{k'}= 15^{\circ}, 30^{\circ} \text{ and } 60^{\circ}$. At $\theta_{k'}= 120^{\circ}$, the normalization of the cross section goes up for this case. For the CBF calculations, a lowering of the strength of the curve with a lower axial mass is seen for both neutrinos and anti-neutrinos at $\theta_{k'}= 60^{\circ}, 120^{\circ}$. In these cases, for more forward angles ($\theta_{k'}= 60^{\circ},120^{\circ}$) this change in the axial mass results in a larger overall cross section normalization.

Pauli blocking, which is included in the nominal for NEUT v3.5.6, was turned off for the purpose of this work. Changing this parameter is expected to only impact the RFG simulations where the momentum distribution has a hard cap for allowed values. Indeed, we see that curves are only impacted when using the RFG model and not the CBF model. The effect of Pauli blocking is only seen for $\theta_{k'}= 15^{\circ}$ in neutrino and anti-neutrino scattering. Removing the Pauli blocking feature results in an increase in the strength of the cross section at lower values of energy transfer for both cases. 

The average binding energy is only included in the RFG model, so it should only have an impact on those cross sections. Again, what is expected physically shows up in the simulations, as the CBF model with the change is consistent with the nominal NEUT v3.5.6. At all angles for both neutrinos and anti-neutrinos, lowering the average binding energy shifts the strength of the curve to lower values of momentum transfer. This effect is anticipated, as the peak in energy transfer should get lower with a lower average binding energy parameter. While the bin sizes are too large to resolve a change in peak position, the increase in strength for bins of lower energy transfer and increase in strength for bins at higher energy transfer demonstrates the occurrence of this change. 

Finally, a dipole vector form factor was used instead of the BBBA05 \cite{bradford2006}. Changing this parameter appeared to only have a significant impact in the RFG model for both neutrinos and anti-neutrinos. In the case of neutrino scattering, changing this feature resulted in an increase in the cross section strength at $\theta_{k'}= 15^{\circ}, 30^{\circ}$ and a decrease in the cross section strength at $\theta_{k'}= 60^{\circ}, 120^{\circ}$. For anti-neutrino scattering, the normalization of the curve is larger at all angles. The CBF model simulations appear to be consistent with the nominal for all angles.

\section{Additional tables and figures}
\label{additional}

\begin{figure*}[ht]
\begin{center}
\includegraphics[width=0.3\textwidth]{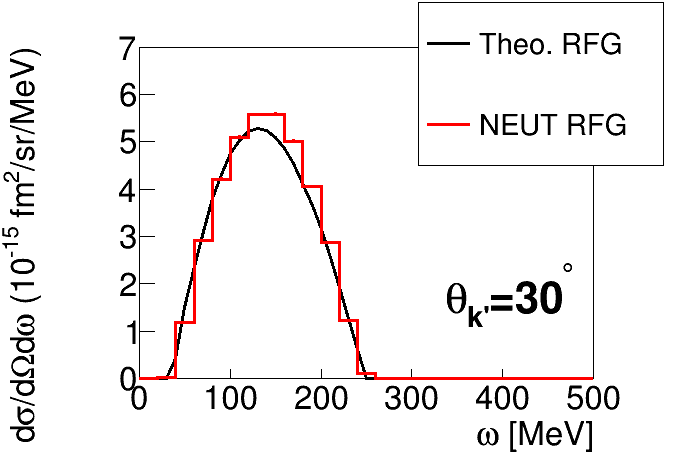}
\includegraphics[width=0.3\textwidth]{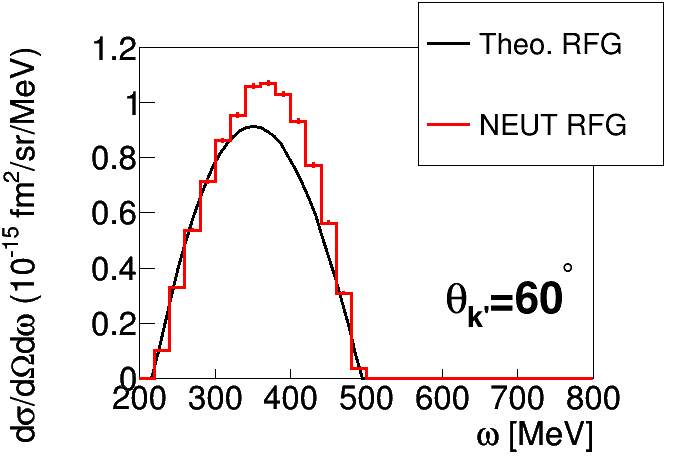}\\
\includegraphics[width=0.3\textwidth]{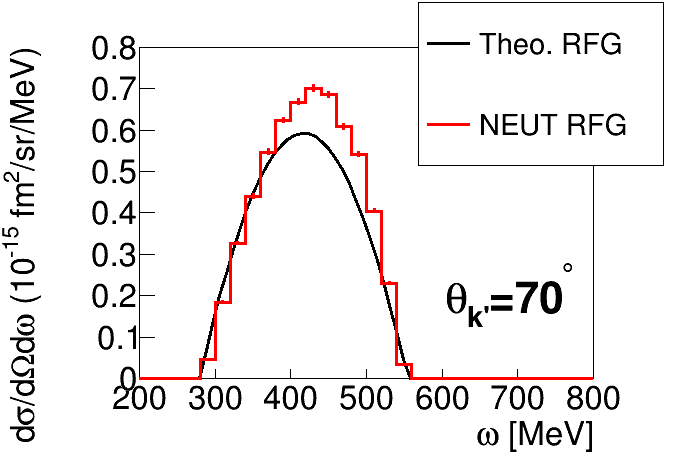}
\includegraphics[width=0.3\textwidth]{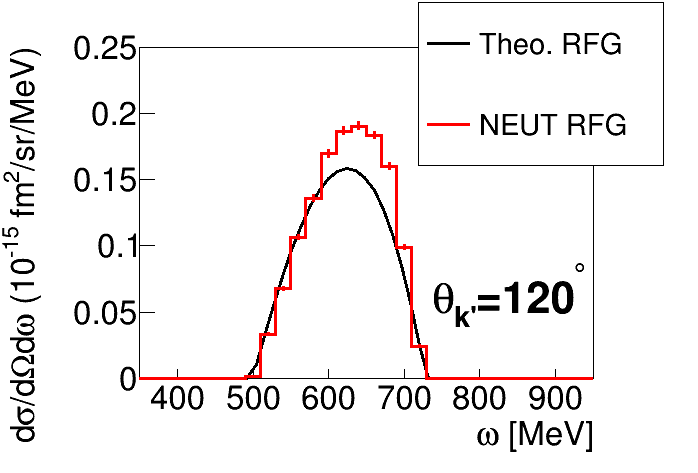}
\end{center}
\caption{Comparison of neutral current quasi-elastic $\nu$-$^{12}$C between cross sections simulated with NEUT  (red) and theory (black) using the RFG model. These comparisons are done at a fixed energy of 1 GeV. The histograms in $\omega$ are binned with widths $\Delta\omega$ = 20 MeV}
\label{fig:nurf}
\end{figure*}

\begin{figure*}[hb]
\begin{center}
\includegraphics[width=0.3\textwidth]{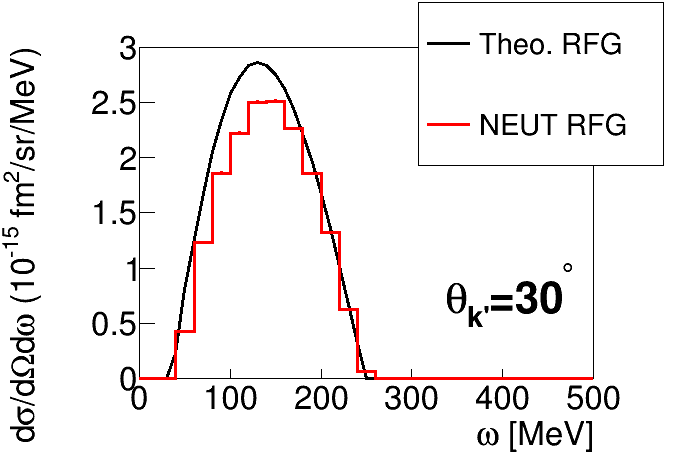}
\includegraphics[width=0.3\textwidth]{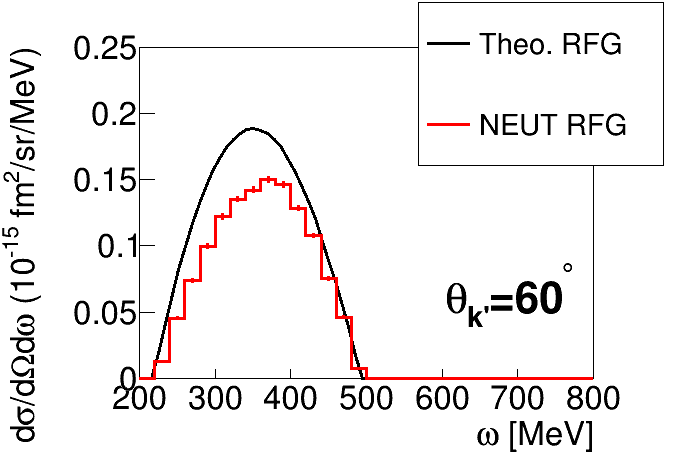}\\
\includegraphics[width=0.3\textwidth]{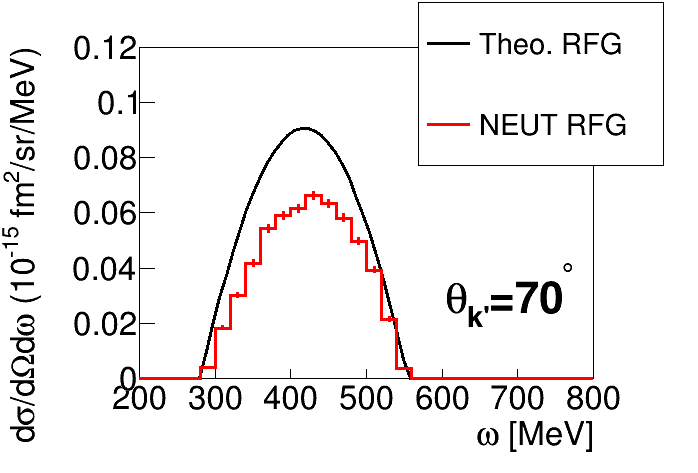}
\includegraphics[width=0.3\textwidth]{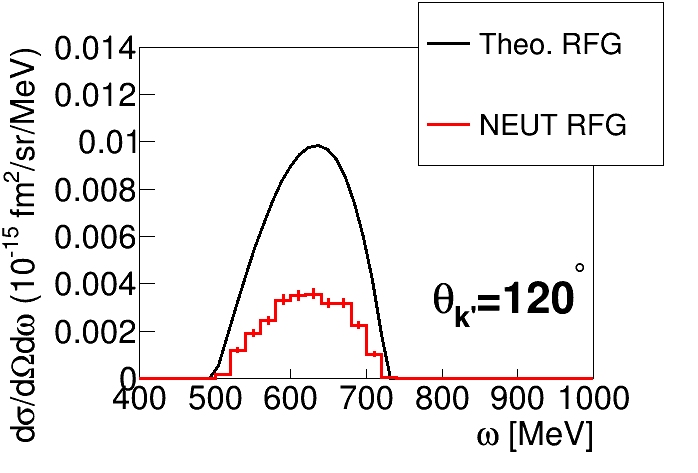}
\end{center}
\caption{ Same as Fig.~\ref{fig:nurf} but for $\bar{\nu}$-$^{12}$C.}
\label{fig:Anurf}
\end{figure*}

\begin{figure*}[ht]
\begin{center}
\includegraphics[width=0.3\textwidth]{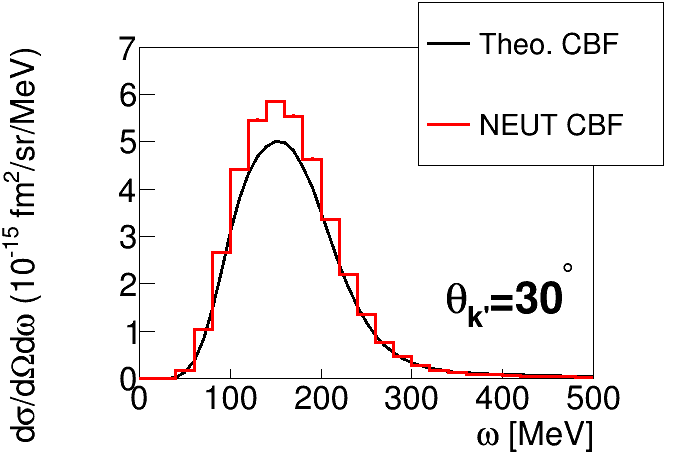}
\includegraphics[width=0.3\textwidth]{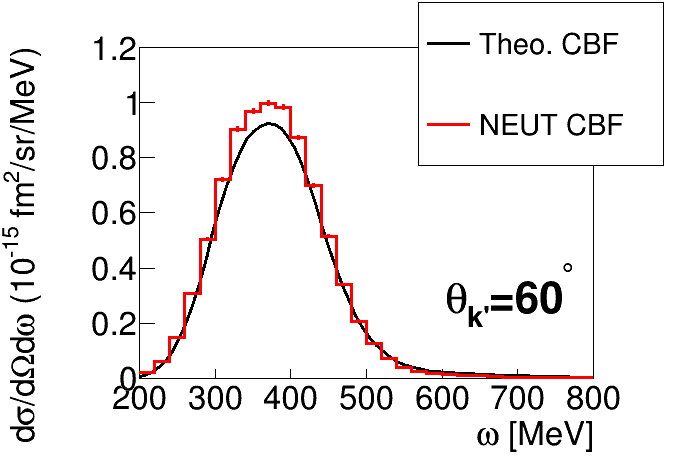}\\
\includegraphics[width=0.3\textwidth]{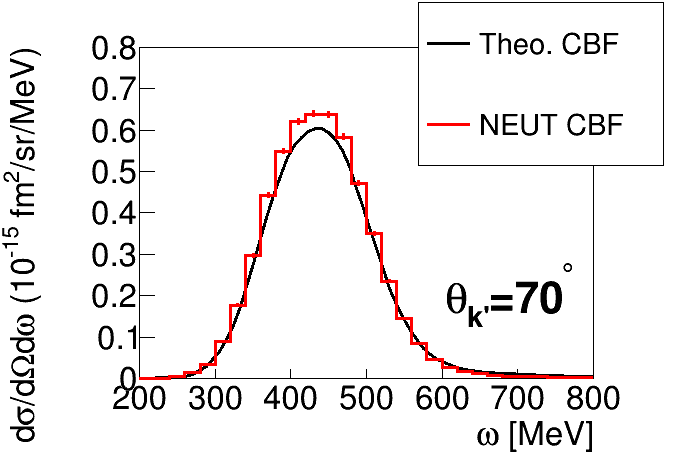}
\includegraphics[width=0.3\textwidth]{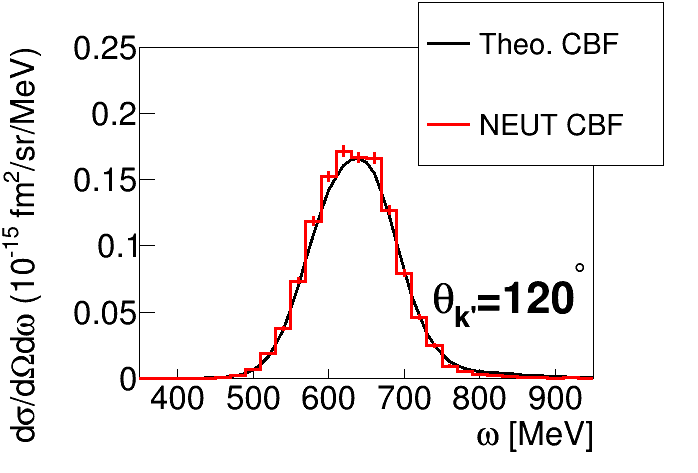}
\end{center}
\caption{Comparison of neutral current quasi-elastic $\nu$-$^{12}$C between cross sections simulated with NEUT  (red) and theory (black) using the CBF model. These comparisons are done at a fixed energy of 1 GeV. The histograms in $\omega$ are binned with widths $\Delta\omega$ = 20 MeV}
\label{fig:nusf}
\end{figure*}

\begin{figure*}[hb]
\begin{center}
\includegraphics[width=0.3\textwidth]{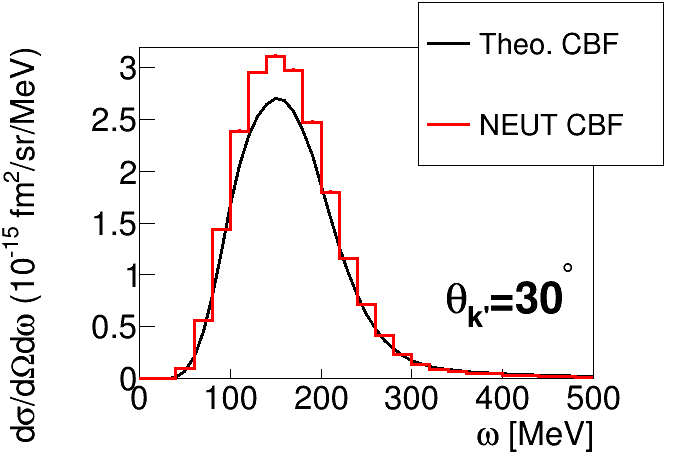}
\includegraphics[width=0.3\textwidth]{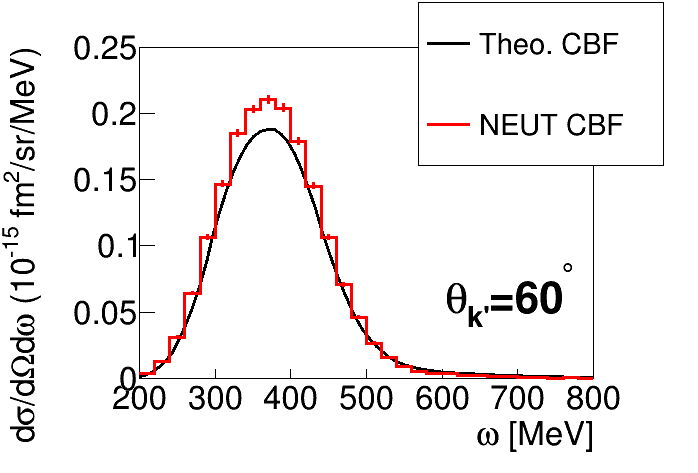}\\
\includegraphics[width=0.3\textwidth]{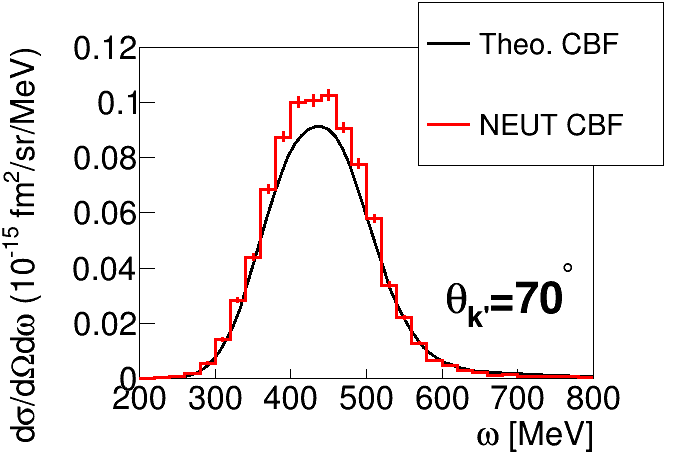}
\includegraphics[width=0.3\textwidth]{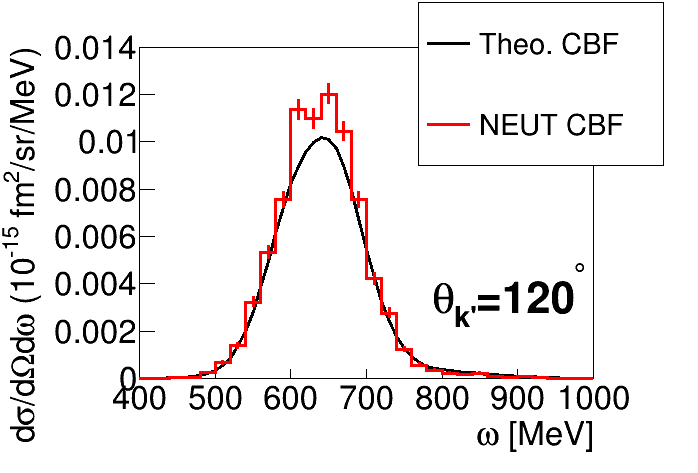}
\end{center}
\caption{Same as Fig.~\ref{fig:nusf} but for $\bar{\nu}$-$^{12}$C}
\label{fig:Anusf}
\end{figure*}

In this section, additoinal tables and figures used in this analysis of EG performance for simulating neutral current quasi-elastic scattering are included. Tables \ref{tab:max-nu} to \ref{tab:tail-Anu} summarize the important shape features of the distributions.  Plots of the double differential cross sections being analyzed are included in Figures \ref{fig:nurf} to \ref{fig:Anusf}. The Tables \ref{tab:transfer-nu} and \ref{tab:transfer-Anu} summarize where in energy transfer $\omega$ the double differential cross section $d\sigma/d\Omega d\omega$ is peaked. The values for NEUT are given as the center of the peak bin in the histogram. The width of the bins is 20 MeV. In all cases, the peak bin in the NEUT simulation overlaps with the peak of the analytic calculation.

\begin{table}[h]
\begin{center}
\begin{tabular}{|c|c|c|c|} \hline
Model &$\theta$ (deg) &  $\omega_{\rm{peak}}^{\text{NEUT}}$ & $\omega_{\rm{peak}}^{\text{Theo}}$ \\ \hline  
\multirow{5}{*}{RFG} &15 & 70 & 70 \\ \cline{2-4}
&30 & 150 & 150 \\ \cline{2-4}
&60 & 370 & 369 \\ \cline{2-4}
&70 & 430 & 432 \\ \cline{2-4}
&120 & 630 & 640 \\ \hline
\multirow{5}{*}{CBF} &15 & 50 & 49 \\ \cline{2-4}
&30 & 150 & 130 \\ \cline{2-4}
&60 & 370 & 351 \\ \cline{2-4}
&70 & 430 &414 \\ \cline{2-4}
&120 & 630 & 624 \\ \hline
\end{tabular}
\end{center}
\caption{The value in energy transfer $\omega$ where $d\sigma/d\Omega d\omega$ peaks for $\nu$-$^{12}$C NCQE scattering at various angles. All values of energy transfer are given in MeV. The NEUT values are the peak bin centers with bin width 20 MeV.}
\label{tab:transfer-nu} 
\end{table}

\begin{table}[h]
\begin{center}
\begin{tabular}{|c|c|c|c|} \hline
Model &$\theta$ (deg) &  $\omega_{\rm{peak}}^{\text{NEUT}}$ & $\omega_{\rm{peak}}^{\text{Theo}}$ \\ \hline  
\multirow{5}{*}{RFG} &15 & 70 & 70  \\ \cline{2-4}
&30 & 150 & 150 \\ \cline{2-4}
&60 & 370 & 369 \\ \cline{2-4}
&70 & 450 & 441 \\ \cline{2-4}
&120 & 650 & 640 \\ \hline
\multirow{5}{*}{CBF} &15 & 50 & 49  \\ \cline{2-4}
&30 & 150 & 130 \\ \cline{2-4}
&60 & 370 & 351 \\ \cline{2-4}
&70 & 430 &414 \\ \cline{2-4}
&120 & 630 & 636 \\ \hline
\end{tabular}
\end{center}
\caption{The value in energy transfer $\omega$ where $d\sigma/d\Omega d\omega$ peaks for $\bar{\nu}$-$^{12}$C NCQE scattering at various angles. All values of energy transfer are given in MeV. The NEUT values are the peak bin centers with bin width 20 MeV.}
\label{tab:transfer-Anu} 
\end{table}

\end{appendix}

\end{document}